\documentclass[referee]{aa}

\usepackage{color}
\usepackage{epsfig}
\usepackage{amssymb}
\usepackage{graphicx}

\begin{document}

\title{New results on source and diffusion spectral features of
Galactic cosmic rays: I- B/C ratio}
         \author{D. Maurin
         \inst{1}
         \and R. Taillet
         \inst{1,2}
         \and F. Donato
         \inst{3}
         }

         \authorrunning{Maurin, Taillet \& Donato}
         \titlerunning{Source and diffusion spectral features\dots}
         \institute{Laboratoire de Physique Th\'eorique {\sc lapth},
         Annecy--le--Vieux, 74941, France
         \and Universit\'e de Savoie, Chamb\'ery, 73011, France
	 \and Universit\`a degli Studi di Torino and INFN, 
	 Torino, Italy
}

\date{Received 18 June 2002; accepted 29 July 2002}

\abstract{
In a previous study (Maurin et al., 2001), we explored 
the set of parameters describing diffusive propagation of cosmic rays 
(galactic convection, reacceleration, halo thickness, spectral index and 
normalization of the diffusion coefficient), and we identified those 
giving a good fit to the measured B/C ratio.
This study is now extended to  take into 
account a sixth free parameter, namely the spectral index of sources.
We use an updated version of our code where the reacceleration
term comes from standard minimal reacceleration models.
The goal of this paper is to present a general view of the evolution
of the goodness of fit to B/C data with the propagation parameters.
In particular, we find that, unlike the well accepted 
picture, and in accordance with
our previous study, a Kolmogorov-like power spectrum for diffusion
is strongly disfavored. Rather, the $\chi^2$ analysis points towards 
$\delta\gtrsim 0.7$ along with source spectra 
index~$\lesssim 2.0$.
Two distinct energy dependences are used for the source spectra: 
the usual power-law in rigidity and a law modified
at low energy, the second choice being only slightly preferred.
We also show that the results are not much affected by a different 
choice for the diffusion scheme.
Finally, we compare our findings to recent works, using other propagation models.
This study will be further refined in a
companion paper, focusing on the fluxes of cosmic ray nuclei.
}

\maketitle


\section{Introduction}
Cosmic rays detected on Earth with kinetic energies per nucleon
from 100 MeV/nuc to 100 GeV/nuc were most probably
produced by the acceleration of a low energy galactic population of
nuclei, followed by diffusion in the turbulent magnetic field.
The acceleration process and the diffusion process have a magnetic
origin, so that they should depend on rigidity.
The rigidity dependence of the diffusion coefficient is given by
quasi-linear theory as
\begin{equation}
    K({\cal R}) = K_0 \beta  \left( \frac{\cal R}{\mbox{1 GV}} \right)^\delta
\end{equation}
where the parameters $K_0$ and $\delta$ should ideally be given by the
small-scale structure of the magnetic field responsible for the
diffusion.  As this structure is not well observed, some
theoretical assumptions must be made in order to predict $\delta$.
As regards the spectrum just after acceleration, the situation is far
from clear, as it depends on the details of the acceleration process.
Several models give a power-law distribution 
({\em e.g.} Berezhko et al., 1994, Gieseler et al., 2000)
\begin{equation}
    \frac{dQ}{dp} \propto {\cal R}^{-\alpha}
\end{equation}
with a definite value for $\alpha$ which depends on the model.

Most analyses of cosmic ray nuclei data assume given power-laws for 
the diffusion and acceleration energy dependence , so that the
results partially reflect certain theoretical {\em a priori}.
In this work, we try to avoid this bias by determining the quantities
$\alpha$ and $\delta$ directly from the data, in particular B/C, for
reasons exposed below.

The paper is organized as follows. We first recall the main features
of our diffusion model. As a few modifications have been made since previous works,
$\S$\ref{niou} is devoted to their description and justification.
Then, the analysis method is described in $\S$\ref{reunnes} and the 
results are shown and discussed in $\S$\ref{reseultses};
a comparison is eventually made with other similar works in $\S$\ref{compar}.

\section{Description of the model}

This paper and its companion (Donato, Maurin \& Taillet, in preparation) use
the same description of cosmic ray propagation as our previous analyses
(\cite{PaperI, PaperII,PaperIII,PaperIV,PaperV}).
Particles are accelerated in a thin galactic disk, from which they
diffuse in a larger volume. When they cross the disk, they may
interact with interstellar matter, which leads to nuclear reactions
(spallations) -- changing their elemental and isotopic composition -- and
to energy losses. Interaction with Alfv\'en waves in the disk also
leads to diffusive reacceleration.
The reader is referred  to Maurin et al. (2001) -- hereafter 
Paper~I -- for  all details, {\em i.e.} geometry and 
solutions of our two zone/three-dimensional diffusive model, nuclear parameters (nuclear grid
and cross sections), energy losses terms (adiabatic, ionization and
Coulomb losses), solar modulation scheme (force-field),
as well as general description of the procedure involved in our
fits to data (selection of a set of parameters, $\chi^2$ test comparison
to data). In particular, though some inputs are modified (see
next section), the final equation describing cosmic ray equilibrium
is formally equivalent to that of Paper~I (see Eq.~A13):
it is a second order differential equation in energy solved with
the Crank-Nicholson approach (see~\cite{PaperII}, Appendix~B -- hereafter 
Paper II).
Finally, a schematic view of our diffusion model is presented
in~\cite{PaperIV} and Fig.~\ref{cascade} (see next section) summarizes
the algorithm of our propagation code.

Some aspects of this model are formally unrealistic.
First, the distribution of interstellar matter has a very simple
structure: it does not take into account a possible $z$ distribution
inside the disk (thin disc approximation is used instead), nor radial and
angular dependence in the galactic plane.
The orthoradial $\theta$ dependence would even be more important from an
accurate description of the magnetic fields and the ensuing diffusion,
as flux tubes are likely to be present along the spiral arms.
However, this is not crucial as we are interested in
effective quantities (diffusion coefficient and interstellar density)
but not in giving them a ``microscopic" explanation.
This is why we chose to use a universal
form of the diffusion coefficient, with the same value in the whole
Galaxy.
Finally, it is known that a fully realistic model has to take into account
interactions between cosmic ray pressure, gas and magnetic pressure,
{\em i.e.} magnetohydrodynamics.

The semi-analytical diffusion approach should be thought of as an intermediate
step between leaky box approaches and magnetohydrodynamics
simulations and is actually justified by these two very approaches:
the first showed that the local abundances of charged nuclei can be roughly
described by two phenomenological coefficients -- the escape length
and the interstellar gas density in the box.
The second hints at the fact that 
the propagation models such as the one used here are well suited
for the description of cosmic ray physics.

However, it is difficult to conclude whether these parameters are valid for 
other kinds of cosmic rays ($e^+$, $e^-$, nuclei induced $\gamma$-ray production)
and whether they are either meaningful but valid
only locally on a few kpc scale ({\em i.e.} not in the whole Galaxy --
see as an illustration Breitschwerdt, Dogiel \& V\"olk (2002)),
or meaningless but phenomenologically valid as an average
description of more subtle phenomena (see as an example the
discussion of the Alfv\'enic speed in Sec.~\ref{ouinde}).


\section{New settings}
\label{niou}
Only a few ingredients differ from our previous analysis (Paper~I). The reason
for these few changes is twofold: first, we attempt to use a better motivated
form of the reacceleration term; second, as the real value of the exponent
in the source power-law cannot be firmly established from acceleration
models -- the latter being seemingly different from what is naively
deduced from direct spectra measurements --, it becomes a free parameter
in the present analysis.


\subsection{Transport of cosmic rays}
\label{transport}
The starting point of all cosmic ray data analysis is the transport equation.
As emphasized in Berezinskii et al. (1990), a diffusion-like
equation was first obtained phenomenologically. 
Afterwards, the kinetic theory
approach provided grounds for a consistent derivation.
This transport equation reads:
\begin{equation}
   \frac{\partial f}{\partial t}-\vec{\nabla} (K\vec{\nabla} f
   -\vec{V_c} f)-
   \frac{\vec{\nabla}.\vec{V_c}}{3}\frac{1}{p^2} \frac{\partial}{\partial p}
   (p^3 f)=
   \frac{1}{p^2}\frac{\partial}{\partial p}p^2 K_{pp}
   \frac{\partial}{\partial p}f +\frac{dQ}{dp}
\end{equation}
In this equation, $f\equiv f(t,\vec{r},\vec{p})$ is the phase space
distribution, $K$ is the spatial diffusion coefficient, $K_{pp}$ is
the momentum diffusion coefficient; 
both are related to the diffusive nature of the process. Finally $V_c$ 
is the velocity describing the convective transport of cosmic rays 
away from the galactic plane.
Actually, the full equation of cosmic ray transport includes 
other terms, such as catastrophic and spontaneous
losses, secondary spallative contributions and continuous energy losses
(coulombian and ionization losses). These were taken into account as 
described in detail in Paper~I, to which the reader is referred for a complete description 
and references. They will not be further discussed here.

This equation can be rewritten using the cosmic ray differential
density $dn/dE\equiv N(E)$. 
As the momentum distribution function is normalized
to the total cosmic ray number density ($n=4\pi \int dp \; p^2f$),
we have $N(E)=(4\pi/\beta) p^2 f$ to finally obtain
\begin{eqnarray}
\label{trans1}
   \frac{\partial N(E)}{\partial t}-\vec{\nabla} \left[
   K\vec{\nabla} N(E) -\vec{V_c} N(E)\right]-
   \frac{(\vec{\nabla}.\vec{V_c})}{3} \frac{\partial}{\partial E}
   (\frac{p^2}{E}N(E))=\\\nonumber
  \frac{\partial}{\partial E}\left[
  -\frac{(1+\beta^2)}{E}K_{pp}\;N(E)
+ \beta^2 K_{pp}\frac{\partial N(E)}{\partial E}
  \right]  + Q(E)\:;
  \label{transport_CR}
\end{eqnarray}
with
\begin{equation}
\label{trans2}
Q(E)\propto\frac{p^2}{\beta}\frac{dQ}{dp}\: .
\end{equation}
In this paper, $K_{pp}$ will be taken from the quasi-linear theory (see below).

From a theoretical point of view, the most natural choice for the 
energy dependence of the source term seems to be a power-law in rigidity 
(or momentum) for $dQ/dp$. 
This translates into $Q(E)\propto R^{-\alpha}/\beta$ in our set of 
equations (see Eqs.~(\ref{trans1}) and~(\ref{trans2}) above). Several different forms
were used in the past because of the lack of strong evidence from
observed spectra (see for example Engelmann et al. (1985), Engelmann
et al. (1990)). In particular, our previous analysis allowed only a rigidity 
dependence $Q(E)\propto R^{-\alpha}$ (for the special case 
$\gamma=\delta+\alpha\approx 2.8$). 
These two forms differ only at low energy and we chose to keep them 
both to estimate their effect on our results. As we show below, it is 
quite small.

Finally, different diffusion schemes lead to different forms for the
energy dependence of the diffusion coefficient and the reacceleration 
term.
Several aspects of the diffusion process are treated in Schlickeiser 
2002, and we considered three alternative possibilities :
(i) Slab Alfven wave turbulence, with 
$K_A(p)=K_0 \beta {\cal R}^\delta$ and $K_{pp}^A\propto V_A^2p^2/K_A(p)$,
(ii) Isotropic fast magnetosonic wave turbulence, with 
$K_F(p)=K_0 \beta^{2-\delta} {\cal R}^\delta$ and $K_{pp}^F\propto V_A^2p^2
\beta^{1-\delta} \ln (v/V_A) /K_F(p)$, 
and (iii) mixture of the two last cases, 
$K_M(p)=K_0 \beta^{1-\delta} {\cal R}^\delta$ and $K_{pp}^M = K_{pp}^F$.

All results will be presented with the case (i), except in the 
specific discussion in Sec.~\ref{discussion_K}


   \subsection{Summary: updates of Paper I's formulae}
   \label{SoC}
The only changes with our previous study are
\begin{itemize}
   \item Eq.~(19) of Paper~I is replaced by
        \begin{equation}
        \label{LOSSES}
                b^j_{loss}(E)=\left< \frac{dE}{dt}\right>_{\rm Ion}+
                        \left< \frac{dE}{dt}\right>_{\rm Coul}
                        +\left< \frac{dE}{dt}\right>_{\rm Adiab}+
                \left< \frac{dE}{dt}\right>_{\rm Reac}\;,
        \end{equation}
        where
        \begin{equation}
        \label{reac1}
        \left<dE/dt\right>_{\rm Reac}=\frac{(1+\beta^2)}{E}K_{pp}\;.
        \end{equation}
   \item Eq.~(A13) of Paper~I (second order differential equation to solve)
        reads now -- we use the same notations --
       \begin{equation}
       A^j_iN^j_i(0)=\bar{{\cal Q}}^j -2h\frac{\partial}{\partial E}
       \left\{b^j_{loss}(E)N^j_i(0)- \beta^2K_{pp}\frac{\partial}{\partial E}
       N^j_i(0)\right\}
       \end{equation}
       with
       \begin{equation}
       K_{pp}= 
\frac{h_{\rm reac}}{h}\times\frac{4}{3\delta(4-\delta^2)(4-\delta)}{V_a}^2p^2/K(E).
       \label{kpp}
       \end{equation}
       In our model, $h_{\rm reac}\equiv h$, but it has to be kept in mind that a possible
       reinterpretation of $V_a$ is always possible (see Sec.~\ref{compar})
       as long as $h_{\rm reac}\ll L$
       (this condition is necessary for the solution to be valid).
   \item As regards the source spectra, two forms
   (hereafter type (a) and (b)) are used instead of Eq.~(9) of Paper I
        \begin{eqnarray}
          \label{Typea}
                a- Q(E) &\propto& \frac{1}{\beta}R^{-\alpha}\\
                \label{Typeb}
                b- Q(E) &\propto& R^{-\alpha}
        \end{eqnarray}
   where $R$ is the rigidity and $\alpha$ a universal slope of
   spectra for all nuclei heavier than helium.
\end{itemize}

\section{Runs and selection method}
\label{reunnes}
The analysis presented here is the natural continuation
of the work presented in Paper~I.
It is more general and it encompasses
its results as a five-dimensional subset of the six-dimensional 
space scanned here.


\subsection{The six free parameters of the study}
\begin{figure}[ht!]
\includegraphics*[width=0.8\textwidth]{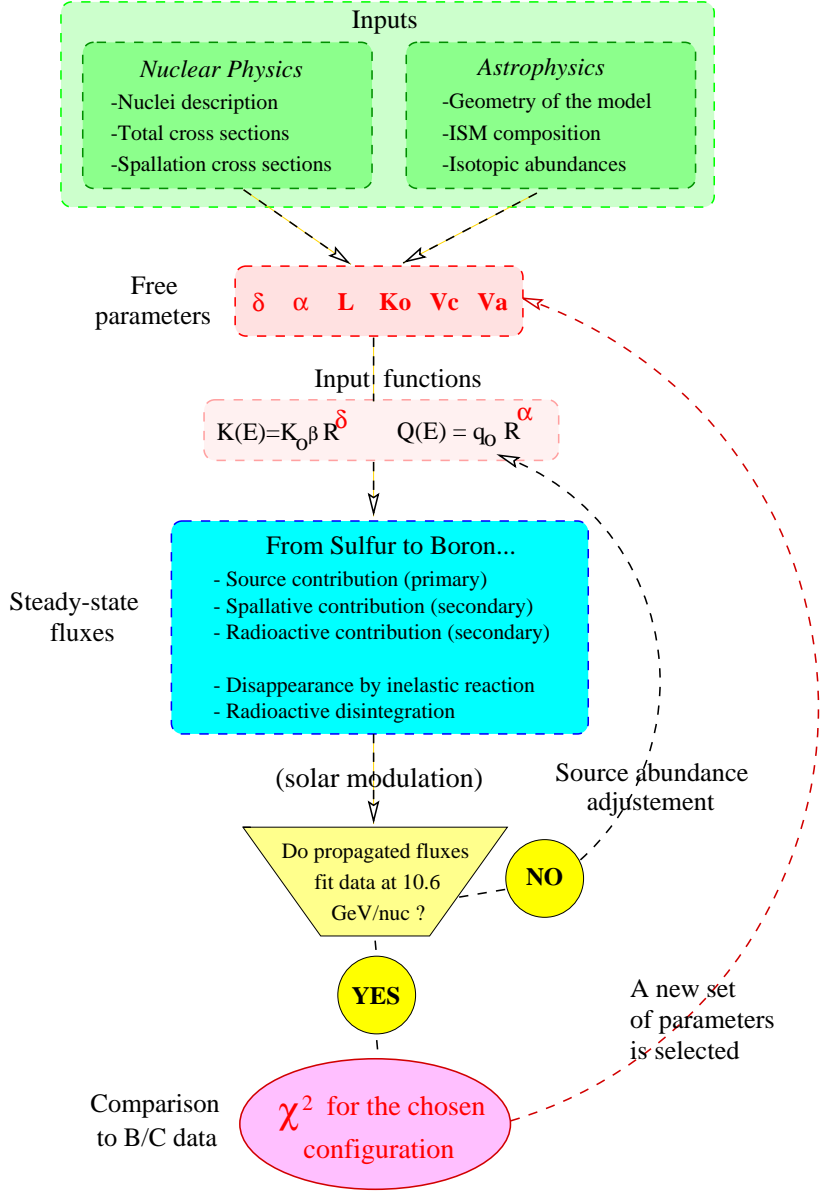}
\caption{Diagrammatic representation of the various steps of the 
propagation code}
\label{cascade}
\end{figure}

The six parameters of this study are: the spectral index of sources
$\alpha$, the normalization $K_0$ and spectral index $\delta$ of the diffusion 
coefficient, the height of the diffusive halo $L$, the Galactic convective 
wind speed $V_c$ and the Alfv\'enic speed $V_a$.
They are included in our code as follows (see Fig.~\ref{cascade} for a
sketch of the procedure): for a given set of parameters,
source abundances of all nuclei ({\em i.e.} primaries and mixed
nuclei) are adjusted so that the propagated
top of atmosphere fluxes agree with  the data at 10.6 GeV/nuc (see Paper~I).
We remind that for B/C ratio, we checked that starting the evaluation of
fluxes from Sulfur is sufficient (heavier nuclei do not contribute
significantly to this ratio).
Top of atmosphere fluxes are deduced from
interstellar fluxes using the force field
modulation scheme (see Paper~I
and references therein).
The resulting B/C spectrum is then compared to the data (see below)
and a $\chi^2$ is computed for the chosen set of parameters.

This procedure is very time consuming. Even when the location of
$\chi^2$ minima in the six-dimensional parameter space are known,
more than $2.10^6$ configurations are needed to have a good sampling
of the regions of interest, for a given form of the source term energy
dependence.

\subsection{$\chi^2$ criterion of goodness}
\label{quideu}
As in our previous analysis, we have computed the quantity
\begin{equation}
    \chi^2 = \sum_i \frac{\left( (B/C)_{i,exp} - (B/C)_{i,model}
    \right)^2}{\sigma_{i, exp}^2}
\end{equation}
where the sum runs over 26 experimental values from {\sc heao}-3 (Engelmann et al., 1990) 
with energies ranging from 620 MeV/nuc to 35 GeV/nuc (as in Paper~I).
In general, if the experimental set-up is such that the measured (experimental)
values differ from the ``real" values by a quantity of zero mean
(non biased) with a given probability distribution, then the value
of $\chi^2$ gives a quantitative estimate of the probability that
the model is appropriate to describe the data.
However, this condition is probably not fulfilled for 
{\sc heao}-3, as for some measured quantity, 
the quoted errors $\sigma_{i, exp}^2$ are much
smaller ({\it e.g.} oxygen fluxes) or much larger ({\it e.g.}
sub-Fe/Fe ratio) than the dispersion of data itself.
For this reason, it is meaningless to associate a likelihood to given
$\chi^2$ values.
Instead, in Paper~I we decided that models giving $\chi^2$ less
than some value $\chi_0^2$ were ``good fits" while the others were
``poor fits".
In this paper, no cut is applied and all the models, whatever the value
of $\chi^2$, are shown in the figures.
\section{Results}
\label{reseultses}


   \subsection{Subset 1: 
fixed measured spectral index $\gamma=\alpha+\delta\equiv2.8$}
\label{arfeuh}
In this section we present the results obtained for source spectra of 
the form $Q(E) \propto R^{-\alpha}/{\beta}$ and diffusion coefficient
$K = K_0 \beta {\cal R}^\delta$.
At sufficiently high energies, spallations and 
energetic changes are irrelevant and the measured fluxes can be 
considered as a mere result of acceleration and diffusion
(see for example Maurin, Cass\'e \& Vangioni-Flam 2002). 
In this case, 
the observed spectrum is proportional to $\cal R^{-\gamma}$ with
$\gamma \equiv \alpha + \delta$.
In this section, we focus on the situation $\gamma = 2.8$, 
corresponding to the spectral index of the measured 
Boron progenitor fluxes. 
Actually, Wiebel--Sooth, Biermann \& Meyer (1998) 
analysed data  from several experiments and derived smaller values. 
In Paper I, we found that the Oxygen flux measured by {\sc heao}-3 
would be more compatible with our diffusion model for a higher $\gamma$, 
namely 2.8 instead of 2.68. 
Anyway, we are more interested in the trends in the variation of 
other parameters for a fixed value of $\gamma$ than in this precise 
numerical value.
The other cases will be treated in the following sections.

\begin{figure}[ht!]
\includegraphics*[width=\textwidth]{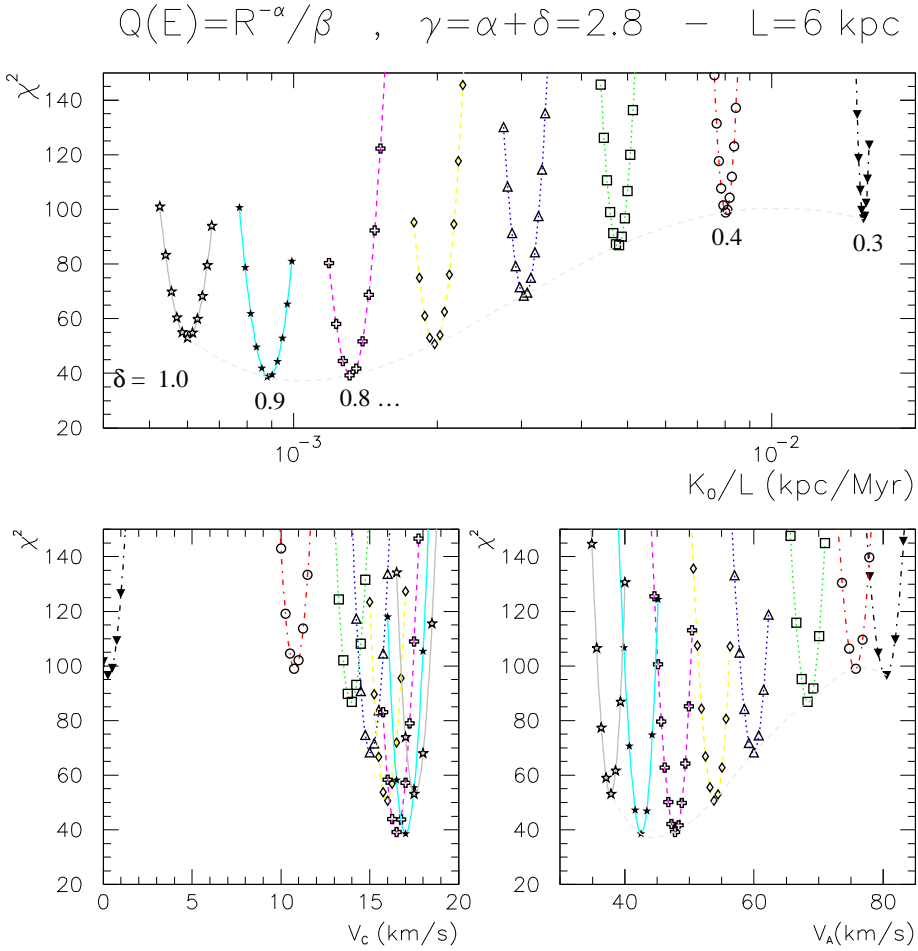}
\caption{Evolution of the $\chi^2$ value for various combination
of parameters.
All curves are for type (a) spectra ($R^{-\alpha}/\beta$) with
$\gamma\equiv\delta+\alpha=2.8$ and the halo size is fixed to $L=6$ kpc.
Each curve shows one-dimensional cuts in the $K_0/L$ (upper panel), $V_c$ (left
lower panel)
or $V_a$ (right lower panel) direction of the 3-dimensional
$\chi^2$ hyper-surface. 
In the upper panel $\delta$ is varied from 1.0 to 0.3 and, in the lower
panels 
the same symbols to indicate $\delta$ are conserved. Each curve gives the 
absolute minimum for the parameter on the abscissa axis, $L$ being fixed to
6 kpc
(similar curves with slightly different minima are obtained for other
$L$ values).}
\label{g28_beta}
\end{figure}

We also set the halo thickness $L$ to 6~kpc, leaving us with
four free parameters ($\delta$, $K_0$, $V_a$ and $V_c$).
All curves depicted in Fig.~\ref{g28_beta} correspond to one-dimensional
cuts through the absolute $\chi^2$ minimum (for a given $\delta$, the 
three different cuts justify the fact that we are located in a minimum).
In the upper panel of Fig.~\ref{g28_beta}, we plot the values of the $\chi^2$
as a function of $K_0/L$, for different values of $\delta$ 
(and the corresponding $\alpha = 2.8 - \delta$).
The best fits are obtained for $\delta \sim$  0.8--0.9,
far from the Kolmogorov spectrum ($\delta$ = 1/3). 
We found a quite similar result in Paper~I, 
where the same assumptions on $\gamma$ were made but with a 
different choice for the source spectrum, $Q(E) = {\cal R}^{-\alpha}$.
The fit is best for values of the diffusion coefficient normalization 
$K_0 \sim 6\times 10^{-3}$ kpc$^2$~Myr$^{-1}$, yielding the value
$\chi^2\sim 40$ (giving $\chi_r^2\sim 1.8$). 
For a Kolmogorov spectrum, the minimum $\chi^2$
is almost twice this value. 
Leaving aside any statistical interpretation of the analysis, we can observe
that for greater $\delta$, the minima of $\chi^2$ are obtained for 
smaller $K_0/L$ (or $K_0$, $L$ being set to 6 kpc) and versa-vice.
This can be understood as at a sufficiently high energy $E_{\rm 
thresh}$, diffusion is the sole remaining influencial parameter and, for
the flux to be unchanged with various $\delta$, one need to satisfy roughly the 
relation $K_0\times E_{\rm thresh}^\delta\approx cte$ (this will also explain
why type (a) and type (b) source spectra give similar $K_0$, see below).

In the lower panels we present two cuts in the two other directions,
namely in the $V_c$ and $V_a$ directions. 
The first one tells us that 
except for the special case $\delta = 0.3$ for which the $\chi^2$ curve 
skips to null $V_c$ , B/C is fitted with $V_c$ between
10 and 20 km~s$^{-1}$. The best $\chi^2$ are for 
convective velocity around $V_c \sim$ 16--18 km~s$^{-1}$. 
For $V_c \lesssim 15$~km~s$^{-1}$ (and $\delta \lesssim 0.6$) the
goodness of the fit quickly decreases. 
We can see that when $\delta$ is around
0.4--0.3, the B/C ratio becomes very sensitive to the $V_c$ values.
It appears that when $\delta$ is 
decreased, a good fit is maintained provided that $V_c$ is also 
lowered. This is possible down to $\delta \sim 0.4$ for which the best 
value for $V_c$ is zero. For lower $\delta$, the previous trade-off 
cannot be achieved (as $V_c$ must be positive for the galactic wind 
to be directed outwards) and no good fit is possible.

The right panel shows the $\chi^2$ curves as functions of the Alfv\'en
velocity.
The minimization procedure always yields a $V_a$ far different from zero. 
Good fits are obtained for values of $V_a \sim$ 40--50 km~s$^{-1}$. 

In each of the explored directions, the $\chi^2$ curves are very narrow: 
the diffusion model leads to meaningful and interpretable 
values for all the physical, free parameters. 
Similar results, with slightly different values for the minima, are
obtained for the other values of $L$ in the range $1\,\leq L \,\leq 15$ kpc.

\begin{figure}[ht!]
\includegraphics*[width=\textwidth]{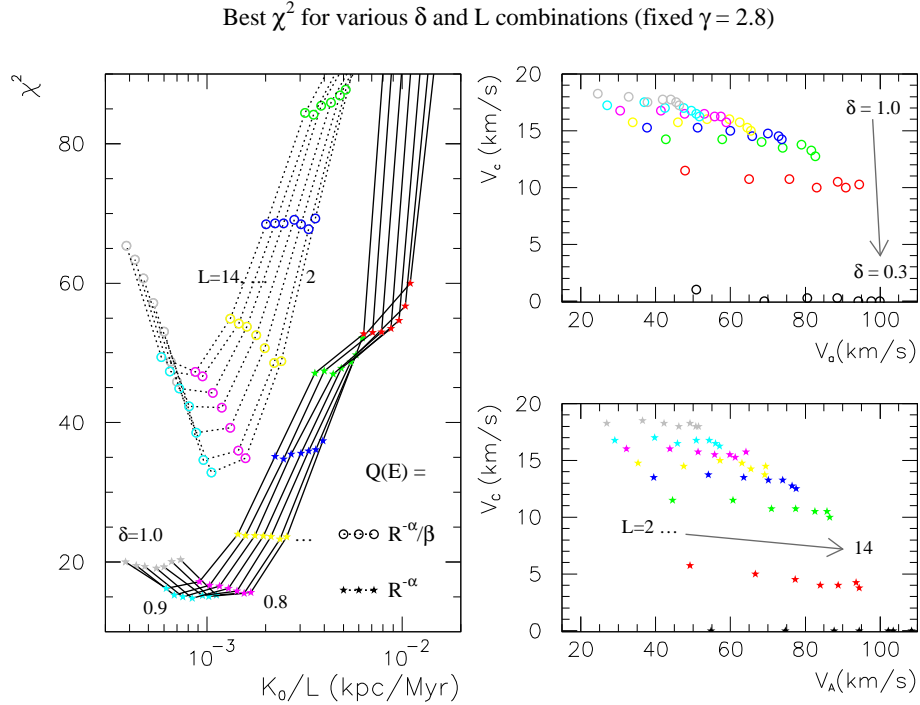}
\caption{Left panel: evolution of the best
$\chi^2$ value with $K_0/L$ for various $\delta$ (1.0 to 0.3, from left to right) at a
fixed $\gamma=2.8$. Each curve correponds to a given halo size $L$ from
14 kpc to 2 kpc. Right panel: the same best $\chi^2$ values are presented 
versus $V_c$ and $V_a$. In both panels, empty circles
correspond to type (a) spectra and stars to type (b) spectra.}
\label{g28_beta_no_beta}
\end{figure}

In Fig.~\ref{g28_beta_no_beta} we present the results for the same analysis
for different values of the halo thickness $L$ and considering also the 
form (b) for the source 
spectra, {\em i.e.} $Q(E) \propto R^{-\alpha}$. 
The total spectral index $\gamma$ is still set to 2.8.
The left panel reports the $\chi^2$ as functions of $K_0/L$, for
different 
values of $\delta$ and $L$, and for both types of source spectra. 
We see that the choice (b) globally improves the fit, and the 
favoured range for $\delta$ is now $\delta \gtrsim 0.4$
(whereas $\delta \gtrsim 0.7$ for choice (a)).
At fixed $\delta$ and $L$, the absolute minima for 
both choices correspond to very similar values of $K_0/L$. 
We can also notice that type (a) spectra 
are, for the higher $\delta$, more sensitive to variations of $L$.

In the right panels we show a cut in the $V_c$--$V_a$ plane. 
For both type (a) and (b) spectra, $\delta$=0.3 
yields a null value for the convective wind.
Type (a) spectra give a little bit higher $V_c$. At fixed $\delta$, the 
variation of $L$  has almost no effect on $V_c$, while it is strongly
correlated with the increase of $V_a$.


   \subsection{Subset 2: $\delta=0.6$, new features from $\alpha$
variation}
\label{deloin}

In this section we discuss the results obtained when the index $\alpha$ 
is varied between 1.3 and 2.5, $\delta$ being set to a given value 
$\delta = 0.6$ which has been extensively used in the literature.

\begin{figure}[ht!]
\includegraphics*[width=\textwidth]{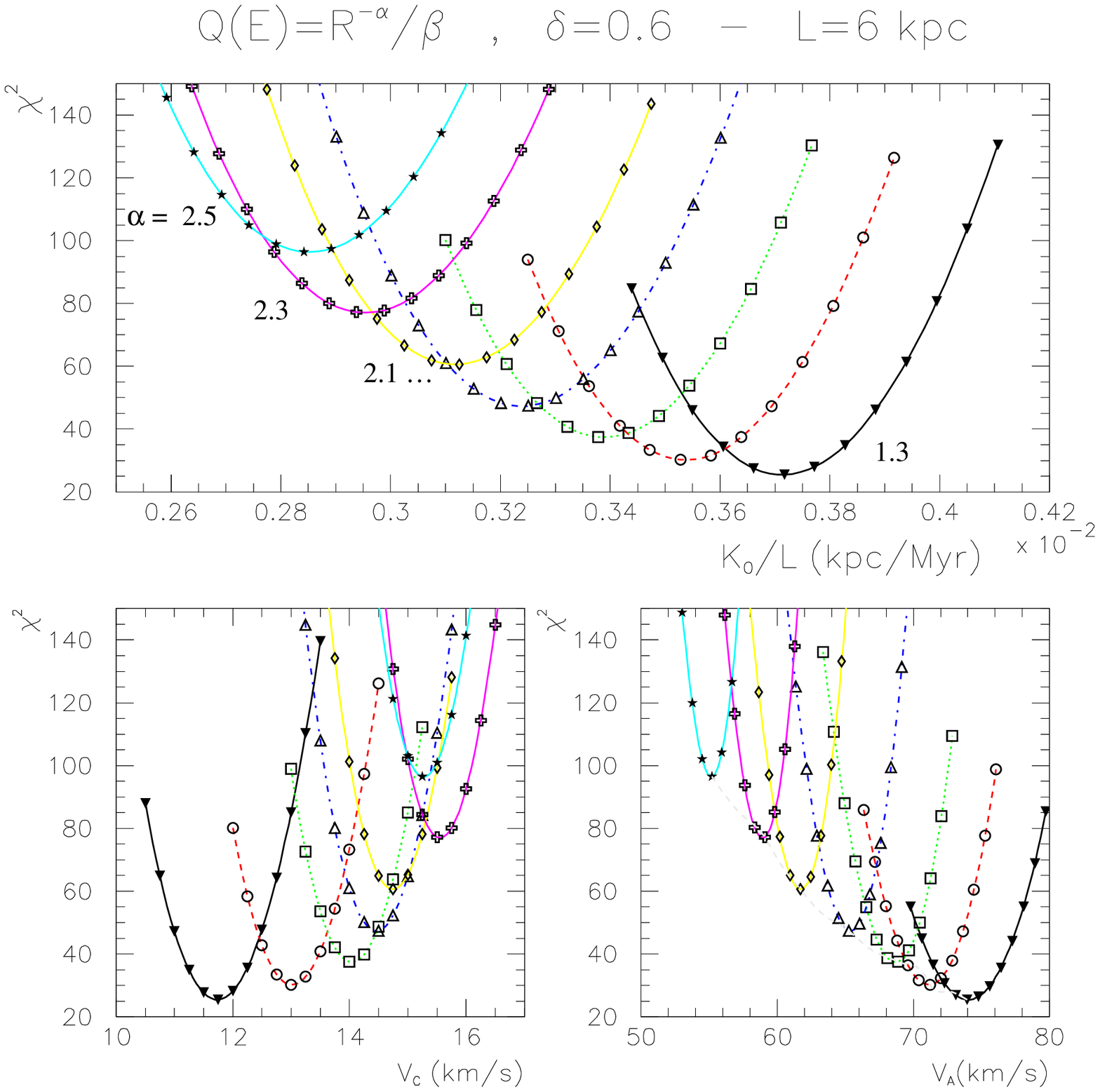}
\caption{Same as in Fig.~\ref{g28_beta} (type (a) spectra, $L=6$ kpc), but
for a fixed $\delta=0.6$.}
\label{d06_beta}
\end{figure}

Fig.~\ref{d06_beta} corresponds to the previous Fig.~\ref{g28_beta}. 
In the left panel we observe that a large variation of the index $\alpha$
has a slight effect on the normalization of the diffusion coefficient 
$K_0$, which stays around an average value $K_0/L \sim 0.0032$ 
kpc~Myr$^{-1}$ for $L=6$ kpc.
Evolution of the absolute $\chi^2$ minimum is also far less
sensitive to $\alpha$ than $\delta$ (see previous section).
However, for $\alpha\gtrsim 2.2$  the fit to the data is poor
and a global power $\gamma \gtrsim 2.8$ at $\delta=0.6$ is excluded.

The lower panels represent a cut in the $V_c$ and $V_a$ directions. 
We can observe that the minimization procedure always drives the minima
towards convective velocities between 12 and 16 km~s$^{-1}$, the least 
$V_c$ being obtained for the smallest $\alpha$. 
This range is again very narrow. 
Similarly, reacceleration is needed to fit data and the minima of the
$\chi^2$ are obtained for $V_a$ between 55 and 75 km~s$^{-1}$. 
Towards this lower limit, $\chi^2$ is high and the
model cannot confidently reproduce observations. 

When $\delta$ is fixed, we can conclude that a variation in the power of
the type (a) source spectrum does not strongly act on the evolution of $K_0,
V_c $ and also $V_a$. 
This can be also easily understood~:
forgetting for a while energy gains and losses, we see from diffusion
equation solutions (the same behavior occurs in leaky box models)
that the source term can be factorized so that secondary to primary ratios
finally do not depend on $Q(E)$, {\em i.e.} are independent of $\alpha$.
Once again, the absolute minimum is identified by a steep $\chi^2$ in
these three directions. 

\begin{figure}[ht!]
\includegraphics*[width=\textwidth]{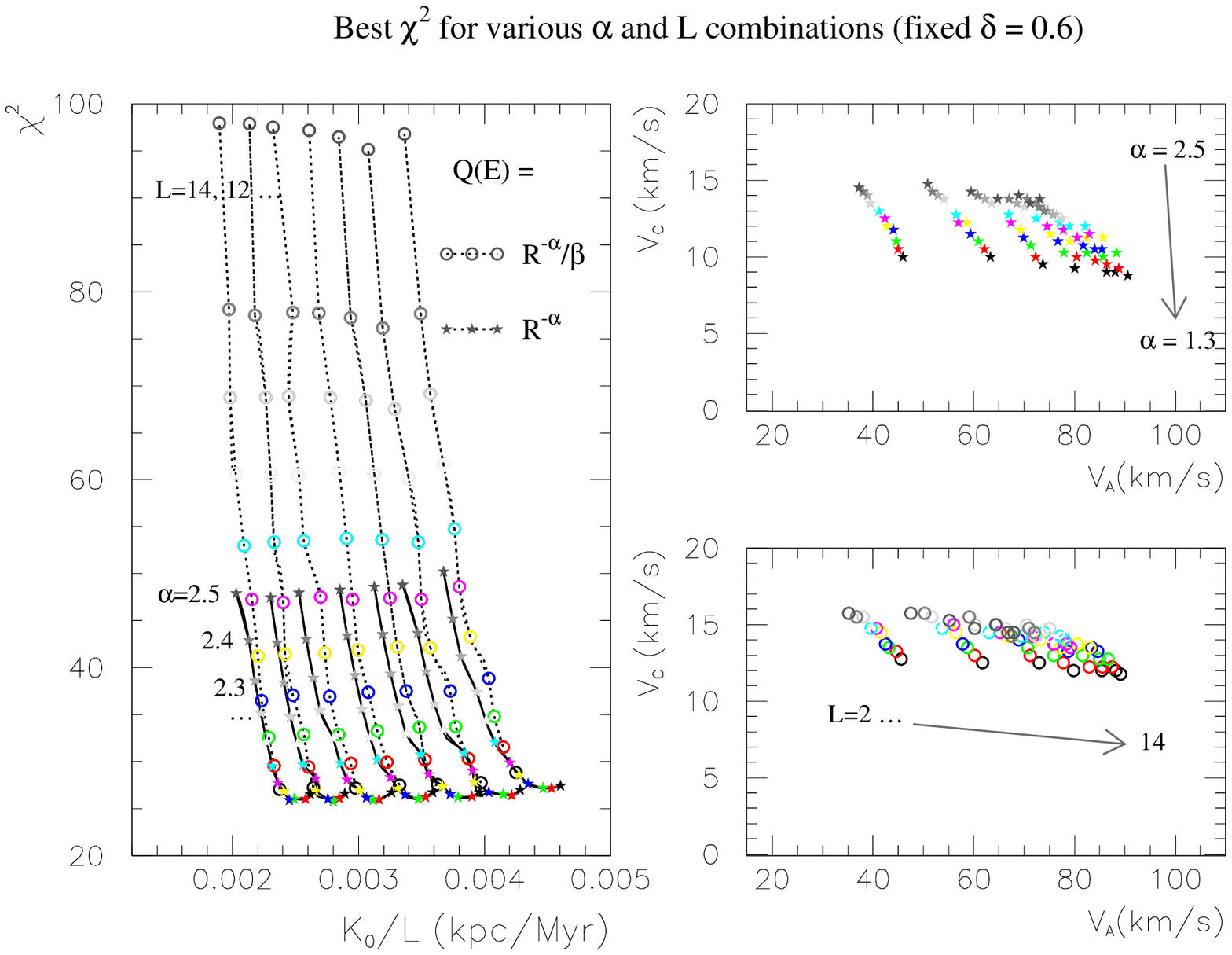}
\caption{Same as in Fig.~\ref{g28_beta_no_beta} but
for a fixed $\delta=0.6$.}
\label{d06_beta_no_beta}
\end{figure}

In Fig.~\ref{d06_beta_no_beta} we present the results for $\delta = 0.6$,
and for both type (a) and (b) source spectra, to focus on the evolution of $L$ and 
$\alpha$.  
The left panel tells us that the evolution of the halo thickness from 2 to
14 kpc, 
at fixed $\alpha$ (in other words, at fixed $\gamma = \alpha + 0.6$)
does not change the goodness of the fit. 
Only a slight modification in $K_0/L$ is required in order to 
recover the same B/C flux ratio. 
Type (b) source spectra reproduce quite well the data for 
all the explored parameter space. On the contrary, the better
theoretically 
motivated type (a) spectra cannot reproduce observations for $\alpha
\gtrsim 2.2$ if $\delta=0.6$.
Since at high energies the two source spectra are equivalent, we must
conclude that it is the low energy part of B/C which is responsible for 
such a discrimination.

The right panels show the absolute minima in the $V_c$--$V_a$ plane. Both
spectra require non--null reacceleration and convection. Even more so, the
selected values reside in the narrow interval for $V_c$, {\em i.e.} 
$V_c\sim 10-15$~km~s$^{-1}$ 
and between 40 and 90 km~s$^{-1}$ for the Alfv\'en velocity.



   \subsection{Subset 3: $\alpha=2.0$, standard acceleration}

\begin{figure}[ht!]
\includegraphics*[width=\textwidth]{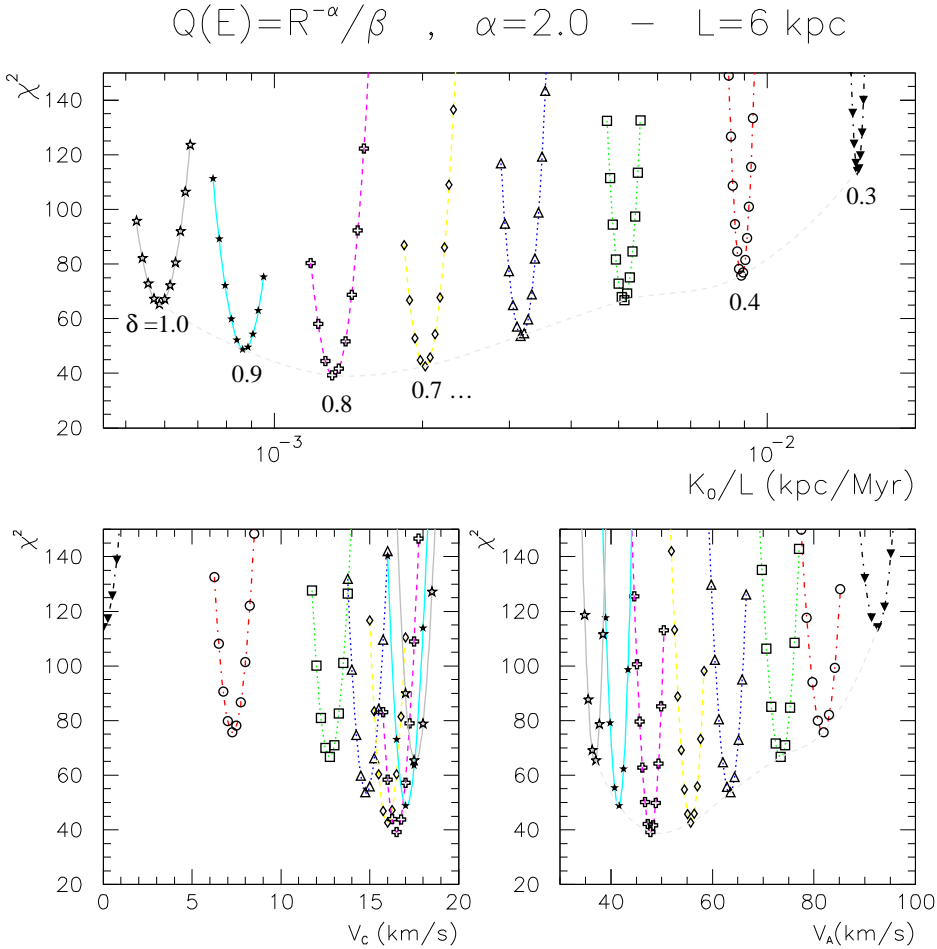}
\caption{Same as in Fig.~\ref{g28_beta} (type (a) spectra, $L=6$ kpc), but
for a fixed $\alpha=2.0$.}
\label{a20_beta}
\end{figure}

Fig.~\ref{a20_beta} describes the results of the analysis done assuming
type (a) 
source spectra, with fixed index $\alpha = 2.0$ and $L=6$ kpc.
A consensus seems to emerge in favor of values $\alpha\approx2.0$
(see Drury et al., 2001 and references therein), close to the index given
by primeval acceleration models, but any other value would 
be fine for the purpose of this section.
In the upper panel $\delta$ has been varied between 1.0 and 0.3, and the
figure 
shows the evolution of the $\chi^2$ with respect to $K_0/L$.
As in Fig.~\ref{g28_beta} and, at variance with Fig.~\ref{d06_beta}, the
minima correspond to $K_0/L$ spanning over almost two orders of magnitude. 
It is the modification of the power-law in the diffusion coefficient --
and not in the source spectrum -- that significantly acts on $K_0$. 
Once again, the Kolmogorov spectrum is disfavoured: in this case
it is obvious that the calculated flux ratio would be too hard.
The best fits are obtained for $\delta \sim0.6$ -- 0.9.

The lower panels show the cuts in the $V_c$ and $V_a$ directions.
The left one tells us that for smaller $\delta$, the preferred convective 
velocities are smaller (and the best $\chi^2$ is larger), down to $\delta = 0.3$ for which a no--convection 
model is prefered, with a bad $\chi^2$.
The best fits are obtained for $V_c$ around 15--18 km~s$^{-1}$.
In the right panel we can notice, again, that only models with reacceleration 
have been chosen by the minimization procedure. 
Lower $\delta$ point to higher $K_0/L$ and $V_a$ values and lower  
$V_c$. The same trend is recovered in the other cases treated above.
Reacceleration and convection act, in a certain sense, in competition,
even if data always give preference to a combined effect rather than 
their absence.

This trend (the smaller $\delta$, the larger $K_0$, or equivalently $K_0/L$ 
as $L$ is constant in the above figures) was already mentioned in 
Sec.~\ref{arfeuh}. Actually, as we will see in Sec.~\ref{compar}, the correlation
between $K_0/L$ and $V_c$ is more properly explained by virtue of 
Eq.~(\ref{gram2}) so that the evolution of $V_c$ is fixed by the evolution
of the two other free parameters, {\em i.e.} $K_0/L$ and $\delta$. 
As regards $V_a$, it only appears in Eq.~(\ref{kpp}). A rough estimation
can be inferred using power-laws $K(E)\propto E^\delta$ and 
$N^j(E)\propto E^{-(\alpha+\delta)}$ in Eqs.~(\ref{trans1}) and~(\ref{kpp}):
\begin{equation}
\frac{d}{dE}\left\{ \frac{V_a^2E}{K_0E^\delta}E^{-(\alpha+\delta)}+
\frac{V_a^2E^2}{K_0E^\delta}(\alpha+\delta)E^{-(\alpha+\delta-1)}\right\}
\label{flan}
\end{equation}
One finally obtains that the term for energetic redistributions evolves 
as $(3+\delta)(1+2\delta)N(E)\times V_a^2/K(E)$ for $\alpha=2$.
Hence, from the above argument, when $\delta$ is decreased,
$K_0$ is adjusted so that $K(E)$ and $N(E)$ remain grossly the same. 
However, for the above expression to be constant, $V_a$ must be increased; 
this is the trend we observe.

\begin{figure}[ht!]
\includegraphics*[width=\textwidth]{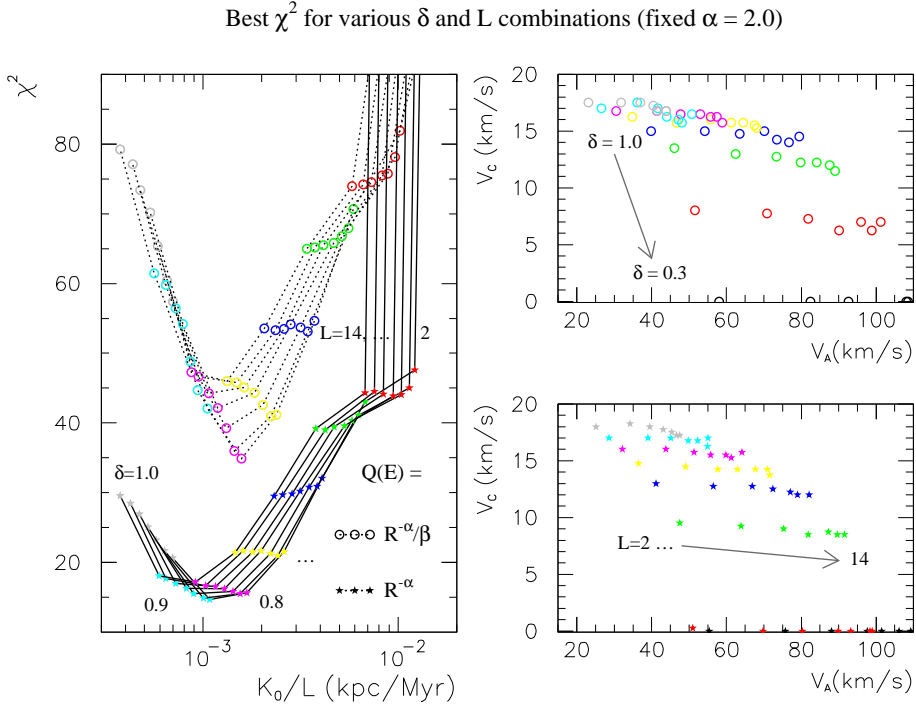}
\caption{Same as in Fig.~\ref{g28_beta_no_beta} but
for a fixed $\alpha=2.0$.}
\label{a20_beta_no_beta}
\end{figure}

In Fig.~\ref{a20_beta_no_beta} we show the effect of varying the halo
thickness when the source spectral index is fixed to 2.0 and all the other 
free parameters are scanned. 
Again, type (b) spectra reproduce better the data. 
When $L$ is varied between 14 and 2 kpc, this may modify the chosen 
$K_0/L$ by a factor of two.
The right panels tell us that the influence of $L$ on $V_a$ is to 
double its value when $L$ is varied from its minimum to its maximum value. 
On the contrary, 
the effect on $V_c$ is almost null. The situation for $V_c$ and $V_a$ is 
very similar to the one discussed in the two above cases, when $\gamma$
and 
then $\alpha$ were fixed. Indeed, looking carefully at the above figures, 
we recover the same effect also for $K_0/L$, at fixed $\alpha + \delta$.
Again, the behaviour of $V_c$ can be understood but cannot
be simply explained. Conversely, neglecting $V_c$ in the asymptotical formula ,
one can see that when $L$ increases, $K_0$ must increase (as can be 
checked in the left panel).
Moreover, it can be seen from the form of $K_{pp}$ that $V_a$ 
increases as the square root of $K_0$ when $\delta$ is fixed 
(see right lower panel).


   \subsection{The whole set: final results}

\begin{figure}[ht!]
\includegraphics*[width=\textwidth]{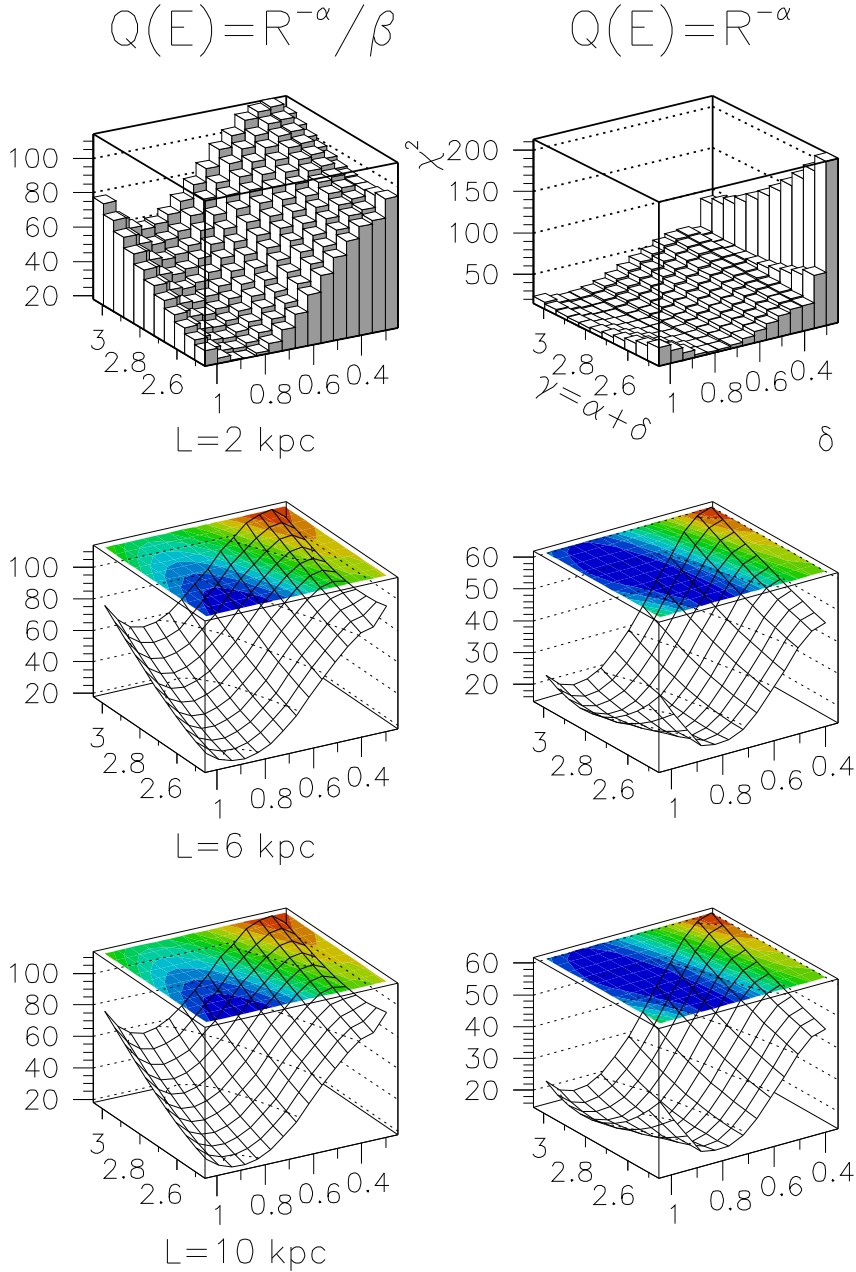}
\caption{Best $\chi^2$ values for various $L$ (2, 6 and 10 kpc) in the
plane $\delta-\gamma$. Left histograms are type (a) spectra and right
histograms type (b). Notice that for right histograms, only the upper figure
displays the values $\delta=0.3$ and $\delta=0.35$. They have been omitted
in the two 
remaining figures
to gain contrast (for any $L$, these configurations have $\chi^2\gtrsim
100$). Assuming $L=6$~kpc, type (a) source spectra give a best value
$\chi^2_{\rm best}=17.8$
for $\alpha=1.65$ and $\delta=0.85$ whereas type (b) gives $\chi^2_{\rm best}=14.6$ for
$\alpha=1.95$ and $\delta=0.85$. These were obtained with 26 data points.
}
\label{final_BC_chi2}
\end{figure}

We know present the result of the full analysis, in which all the 
parameters are varied.
Fig.~\ref{final_BC_chi2} shows the evolution of the $\chi^2$ in the 
$\delta$ and $\gamma = \delta +\alpha$ plane for different values of $L$.
We can see that, at fixed type (a) or (b) spectra, a change in the halo height
$L$ has almost no effect on the best $\chi^2$ surface. 
Generally, high values for $\delta$ are preferred and, a Kolmogorov regime
for the spatial diffusion coefficient is strongly disfavoured over all the
parameter space. 
More precisely, type (b) spectra point towards a band defined
by $\delta \sim 0.8$ in the $\delta$--$\gamma$ plane, whereas the type (a)
spectra gives the additional constraint $\gamma \lesssim 2.8$
(see Fig.~\ref{final_BC_chi2}).

\begin{figure}[ht!]
\includegraphics*[width=\textwidth]{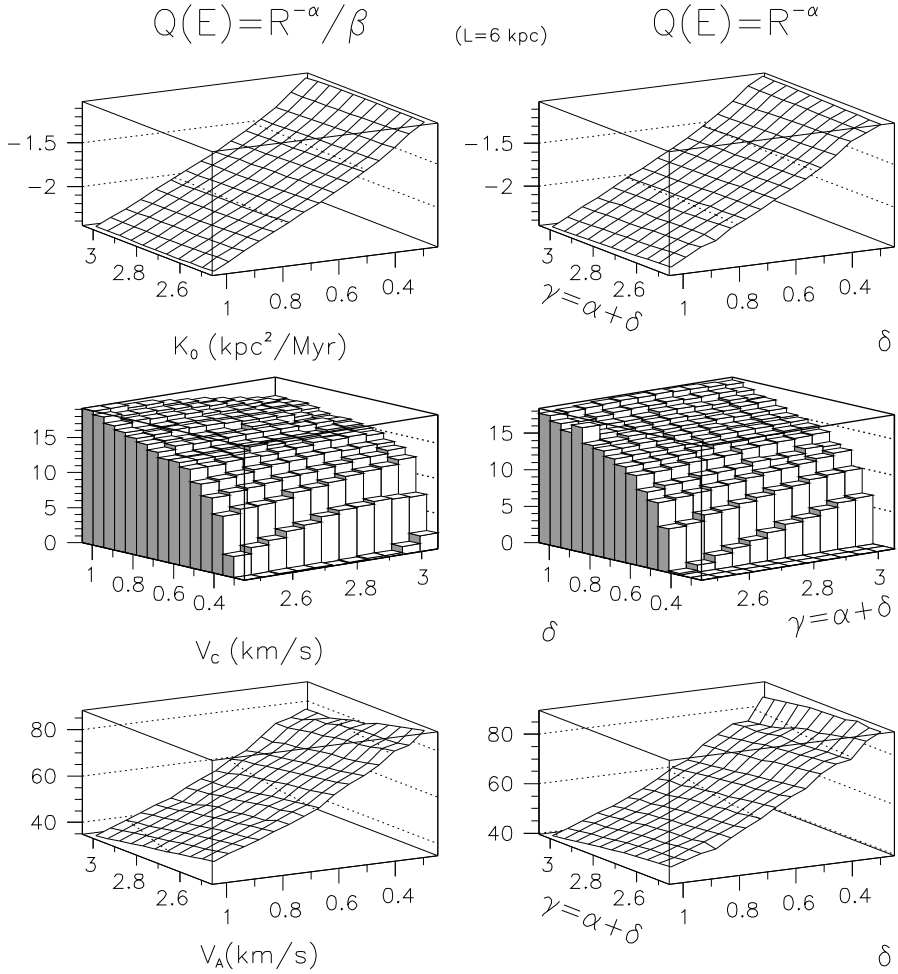}
\caption{From top to bottom: for each best $\chi^2$ in the plane
$\delta-\gamma$
($L=6$ kpc), the corresponding values of $\log(K_0)$, $V_c$ and $V_a$
are plotted for both source spectrum types.}
\label{final_BC_reste}
\end{figure}

In Fig.~\ref{final_BC_reste} we show the preferred values of the three
remaining 
diffusion parameters $K_0$, $V_c$ and $V_a$, for each best $\chi^2$ in the 
$\delta$--$\gamma$ plane, when $L$ has been fixed to 6 kpc.
The two upper panels show that the evolution of $\alpha$ does not affect
$K_0$.
On the other hand, as already noticed, we clearly see the (anti)correlation
between the two parameters $K_0$ and $\delta$ entering the diffusion 
coefficient formula, giving the same normalization at high energy 
($K_0\times E_{\rm thresh}^\delta\approx cte$). 
Almost the same 
numbers are obtained for type (a) and (b) spectra. $K_0$ spans between 
0.003 and 0.1 kpc$^2$~Myr$^{-1}$. We will discuss in the following sections 
how these results can be compared to the literature.

The middle panels show the values for the convective velocity. Only very
few configurations include $V_c=0$, always when $\delta=0.3$, for 
both types of source spectra. The value of $V_c$ increases with $\delta$. 
For type (a) spectra, increasing $\gamma$ and $\delta$ at the same time
makes $V_c$ change its trend.
As remarked previously, the effect of Galactic
wind is more subtle since it acts at intermediate energies
and is correlated with all the other diffusion parameters through
the numerous terms of the diffusion equation.

The lowest two panels show the influence of $V_a$. We recover a
correlation similar to the one discussed for $K_0$ (see Eq.~\ref{flan}).
The Alfv\'en  velocity doubles from $\delta = 1.0$ to 0.3, whereas it 
is almost unchanged by a variation in the parameter $\gamma$ (or equivalently
$\alpha$). 

All the three analysed parameters ({\em i.e.} $K_0$, $V_c$ and $V_a$) 
behave very similarly with respect to a change in the source spectrum 
from type (a) to type (b). It can be explained as the influence
on the primary and secondary fluxes can be factored out (see 
Sec.~\ref{deloin}) if energy changes are discarded (their effect is 
actually small on the derived parameters).
Existing data on B/C do not allow us to discriminate clearly between these two 
shapes for the acceleration spectrum. This goal could be reached by means
of better data not only for B/C but also for primary nuclei (Donato et al., in 
preparation).

\subsection{Other diffusion schemes}
\label{discussion_K}

\begin{figure}[ht!]
\includegraphics*[width=0.7\textwidth]{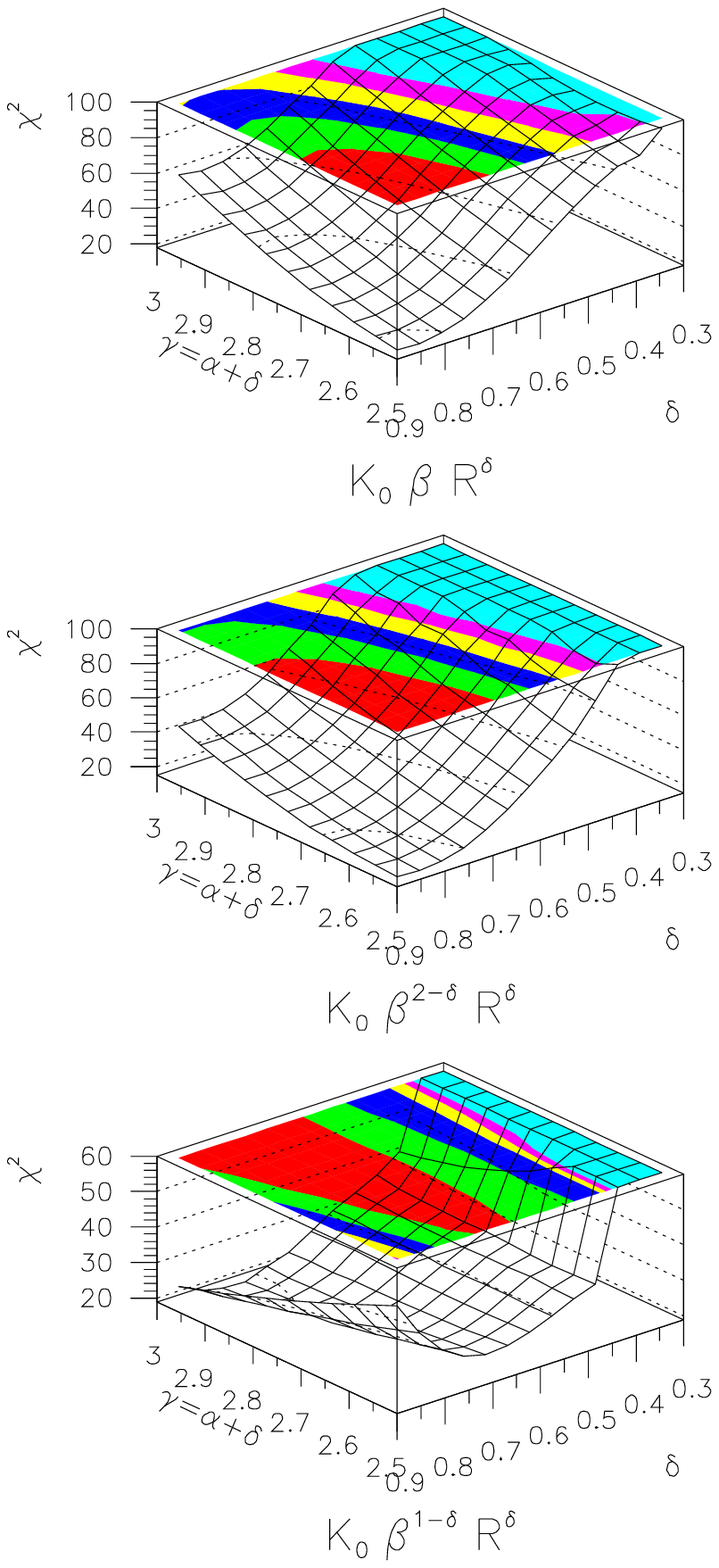}
\caption{Best $\chi^2$ values, in the
plane $\delta-\gamma$, for the three different forms
of the diffusion coefficient and reacceleration terms 
(i) Slab Alfven wave turbulence, with 
$K_A(p)=K_0 \beta {\cal R}^\delta$ and $K_{pp}^A\propto V_A^2p^2/K_A(p)$,
(ii) Isotropic fast magnetosonic wave turbulence, with 
$K_F(p)=K_0 \beta^{2-\delta} {\cal R}^\delta$ and $K_{pp}^F\propto V_A^2p^2
\beta^{1-\delta} \ln (v/V_A) /K_F(p)$, 
and (iii) mixture of the two last cases, 
$K_M(p)=K_0 \beta^{1-\delta} {\cal R}^\delta$ and $K_{pp}^M = K_{pp}^F$.}
\label{tests_K}
\end{figure}

As discussed in Sec.~\ref{transport}, we tested three different 
diffusion schemes, with three different forms for the diffusion coefficient.
Most results are basically insensitive to the choice of this form.
In particular, the figures corresponding to Fig.~\ref{final_BC_reste} 
are almost identical to the case presented above, so that they will 
not be reproduced here.
Figure~\ref{tests_K} displays the $\chi^2$ as a function of $\delta$ and $\gamma$.
The values of $\chi^2$ are slightly different in the three cases, but 
the general trend is the same, and all the previous conclusions still 
apply.

   \subsection{Sub-Fe/Fe ratio}
\label{SubFe}

In an ideal situation in which we had very good and consistent data on B/C and 
sub-Fe/Fe ratios, the best attitude would be to make a statistical 
analysis of the combined set of data. Unfortunately, this is not 
currently the case.
We consider two ways to extract information from the Sub-Fe/Fe data.
First, as a check, we compare the sub-Fe/Fe
ratio predicted by our model -- using the parameters derived
from our above B/C analysis -- with data from the same experiment.
Second, we search directly the minimum $\chi^2_{\rm Fe}$ of
the sub-Fe/Fe ratio, with no prior coming from B/C. 
As previously emphasized (see Sec.~\ref{quideu}), 
this procedure is more hazardous since the
statistical significance of the sub-Fe/Fe data is far from clear.


\subsubsection{Using B/C-induced parameters to derive $\chi^2_{{\rm B/C} \rightarrow 
{\rm Fe}}$}

For each set of diffusion parameters giving a good fit to the observed 
B/C ratio, the sub-Fe/Fe ratio can be computed and compared to the 
values measured by {\sc heao}-3.
This is not as straightforward as in the B/C case because
although Sc, Ti and V -- that enter in the
sub-Fe group (as combined in data here) -- are pure secondaries,
some of the species intermediate between sub-Fe and Fe, contributing 
to the sub-Fe flux, are mixed species ({\em i.e.} Cr, Mn). 
As a consequence, all the primary contributions 
were adjusted so as to reproduce the sub-Fe/Fe ratio at 3.35 GeV/nuc.
The sub-Fe/Fe spectra are not steep enough at high energy, so that 
normalization at 10.6 GeV ({\em i.e.} as for B/C) would have led to 
less good fits. 
We emphasize that to perform this normalization of secondary-to-primary is 
equivalent to making an assumption about the elemental composition of the 
sources, which is usually deduced from secondary-to-primary ratios.
A different choice would slightly shift the normalization of sub-Fe/Fe ratio
without affecting much our conclusions.

\begin{figure}[ht!]
\includegraphics*[width=\textwidth]{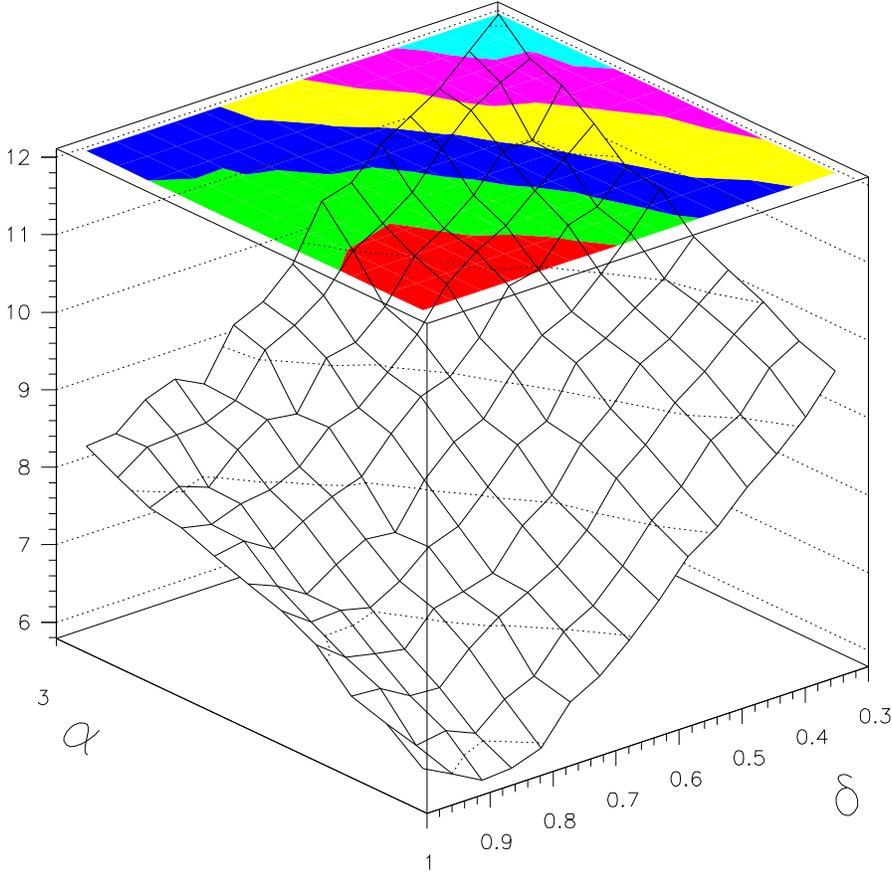}
\caption{Values of $\chi^2_{{\rm B/C} \rightarrow {\rm Fe}}$ 
obtained by applying the diffusion 
parameters -- type (a) source spectra and $L=6$~kpc -- giving the best fit to B/C, for each $\alpha$ and 
$\delta$, to sub-Fe/Fe. The $\chi^2_{{\rm B/C} \rightarrow {\rm Fe}}$ 
are computed with {\sc heao}-3 data points.}
\label{chi2_subfe_BC_values}
\end{figure}

Fig.~\ref{chi2_subfe_BC_values} displays the $\chi^2_{{\rm B/C} \rightarrow 
{\rm Fe}}$  values obtained when the diffusion parameters  
giving a good fit to B/C are used to compute the sub-Fe/Fe ratio ,
for each value of $\alpha$ and $\delta$ (for type (a) spectra 
and $L=6$~kpc,
although the results for type (b) and/or different $L$ are quite
similar).
This surface is very similar to the surface obtained with B/C, 
pointing towards high values of $\delta$ (compare to Fig.~\ref{final_BC_chi2}).


\subsubsection{Looking for $\chi^2_{\rm Fe}$}

\begin{figure}[ht!]
\includegraphics*[width=\textwidth]{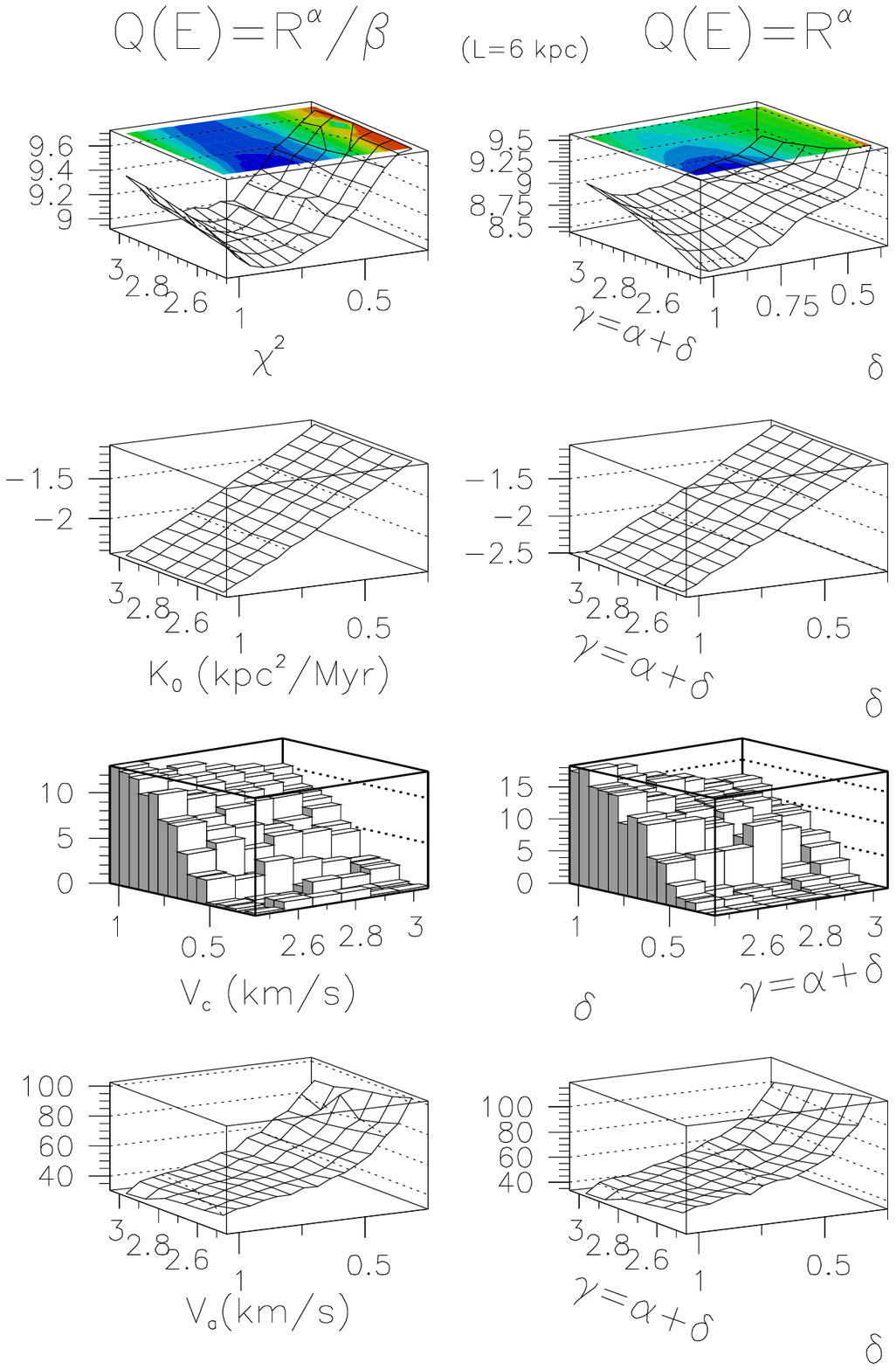}
\caption{From top to bottom: best $\chi^2_{\rm Fe}$ and for each best 
$\chi^2_{\rm Fe}$ in the $\delta-\gamma$ plane 
($L=6$ kpc), the corresponding values of $\log(K_0)$, 
$V_c$ and $V_a$
are plotted for both source spectrum types.}
\label{chi2_subfe}
\end{figure}

We now consider a full sub-Fe/Fe analysis ({\em i.e.} the parameters
minimizing $\chi^2_{\rm Fe}$ are looked for) but we emphasize
that the results given here are from our point of view
far less robust than those obtained from B/C. 
As a consequence, conclusions of this
section have to be taken only as possible trends. 
Several points can be underlined from Fig.~\ref{chi2_subfe}: 
(i) as for the B/C case, the best $\chi^2_{\rm Fe}$
is obtained for type (b) spectra.
(ii) the general behavior of $K_0$, $V_c$ and to a less extent $V_a$ 
is mostly the same as for B/C.
(iii) the type (b) spectra yield propagation parameters which 
are closer to B/C's, as 
we can see from $V_c$ values;  
(iv) finally, consistency with B/C analysis would be better obtained 
for $\delta$ pointing towards $0.6-0.7$.


\subsection{Additional insight from visual comparison of our
model to data}

\begin{figure}
\includegraphics*[width=\textwidth]{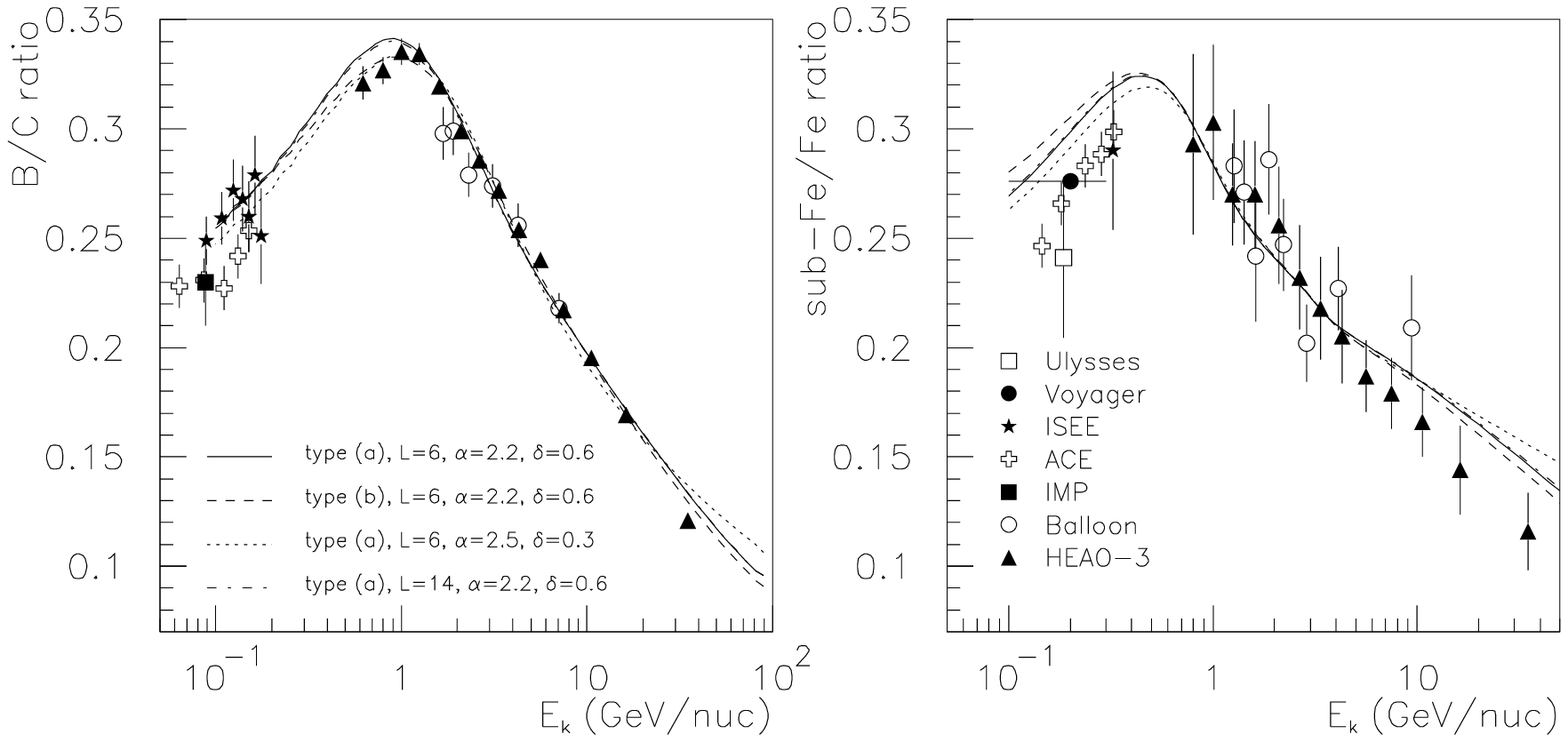}
\caption{The B/C and sub-Fe/Fe spectra (modulated at $\Phi=500$~MV)
for several sets of parameters (giving 
the best fit to B/C for these values) are displayed, along with experimental data 
from {\sc heao}-3 (Engelmann et al., 1990), 
ballon flights (Dwyer \& Meyer, 1987),
{\sc het} on Ulysses (Duvernois \& Thayer, 1996),
{\sc hkh} on {\sc isee}-3 (Leske, 1993)
and Voyager (Webber, Lukasiak \& McDonald, 2002). Note
that {\sc ace} data (Davis et al., 2002) correspond to a modulation parameter
$\Phi\approx750$~MV.}
\label{BC_SubFe}
\end{figure}

Typical spectra (modulated at $\Phi=500$~MV) are shown in Fig.~\ref{BC_SubFe}, for 
different values of the parameters $\alpha$ and $\delta$, along with 
the data points from {\sc heao}-3 (Engelmann et al., 1990) and balloon 
flights (Dwyer \& Meyer, 1987).   
Three low-energy data points, from {\sc het} on Ulysses (Duvernois \& Thayer, 
1996), {\sc hkh} on {\sc isee}-3 (Leske, 1993) and Voyager 
(Webber, Lukasiak \& McDonald, 2002) are also shown; they all have about the
same modulation parameter, {\em i.e.} $\Phi\approx500$~MV.
The {\sc ace} points ($\Phi\approx750$~MV) are also displayed 
(Davis et al., 2002). 

All the models displayed give similar spectra, which would be 
difficult to sort by eye. This may explain why some of these models ({\em e.g.}
those with $\delta=0.3$) are retained in other studies.
The main features are (i) the influence of $\delta$ on the high energy 
behaviour -- a good discrimination between these models would be provided 
by precise measurements around 100 GeV/nuc -- and  (ii) the type (a) 
source spectra are steeper than type (b) at low energy.

\section{Comparison with other works}
\label{compar}
Some of our configurations can be compared to those previously
found in similar models. In particular, to
compare the Alfv\'en speed from one paper to another, we have to be
sure that all $V_a$ used denote the same quantity. 

To compare the reacceleration terms employed, we retain only 
the spallation term and the highest order derivative in energy in
the diffusion equation, giving
\begin{equation}
\label{robert1}
2hn_{\rm ISM}v\sigma_jN_j(0)=2h\beta^2 K_{pp}\frac{\partial^2 N_j(0)}
{\partial E^2}\:.
\end{equation}
We have supposed that both phenomena occur only in the thin disc $h\ll L$ 
and, in the above
equation, the reacceleration zone height equals the spallative zone height.
If it is not the case, we have to correct the previous relation by a
multiplying factor $h_{\rm reac}/h$. Actually, $V_a$ is ``fixed" 
through the choice of $K_{pp}K(E)$.
As underlined in Sec.~\ref{transport}, this paper now follows the
requisites of minimal reacceleration models (see Tab.~\ref{comparaison1}, 
last line).

Once this $h_{\rm reac}/h_{\rm gas}$ rescaling -- that differs from one 
paper to another -- is taken into account, a comparison 
is possible between models if a minimal resemblance exists between the other input
parameters, 
{\em i.e.} same $\delta$, $\alpha$ (plus same form of the source spectrum) and 
halo size $L$; Tab.~\ref{comparaison1} shows the value adopted for  these parameters 
in two recent studies.
\begin{table}[ht]
\caption{Main characteristics of various diffusion models}
\label{comparaison1}
\begin{center}
\begin{tabular}{|c|c|c|c|}   \hline
   & Maurin et al. / This work & Seo \& Ptuskin / Jones et 
al. & Moskalenko et al.\\
   & (2001) / (2002) & (1994) / (2001) & (2002)\\ \hline\hline
Thin disc h (pc) &  $h\equiv100$ pc &  $h=200$ pc& Gas distribution\\
Halo size $L$ (kpc) & --- & $L=3$ kpc & $L=4$ kpc\\
$2h_{\rm reac}$& $2h_{\rm reac}=2h$  & $2h_{\rm reac}=2L/3^{\dagger}$ & $2h_{\rm 
reac}=2L$\\
Surface mass density$^\ddagger$  & $\simeq 10^{-3}$ g cm$^{-2}$&
$\simeq 2.0\times10^{-3}$ / $2.4\times10^{-3}$ g cm$^{-2}$ &
$\simeq 1.6\times10^{-3}$ g cm$^{-2}$ \vspace{0.5mm}\\
$K_{pp}K(E)$ &
$ \frac{2p^2V_a^2}{9}$ /
 $\frac{4p^2V_a^2}{3\delta(4-\delta^2)(4-\delta)w}$
& $\frac{4p^2V_a^2}{3\delta(4-\delta^2)(4-\delta)w}$
& $\frac{4p^2V_a^2}{3\delta(4-\delta^2)(4-\delta)w}$
\vspace{1.mm}\\\hline
\end{tabular}
\end{center}
{~$^\dagger$ Jones et al. use the same set of equation and parameters
as Seo and Ptuskin for the stochastic reacceleration model. Consequently,
it seems that $h_a$ defined in Jones et al. is the half-height of the 
reacceleration zone, contrarily to what is depicted in their Fig.~4.}\\
$^\ddagger$ The surface mass density is defined as $\mu=2h\rho$ where $\rho$ is
the matter density in the thin disk.
\end{table}


        \subsection{Maurin et al. (2001) -- Paper I}

The results are expected to be slightly different from our previous 
study as the components have been modified.
First, $V_a$ has a different interpretation in the two studies
(see Tab.~\ref{comparaison1}, first column).
As underlined above -- remembering that in Paper~I the diffusion coefficients scaled 
as $K_{pp}K(E)\equiv (2/9)\times p^2V_a^2$ --, the Alfv\'en speed value 
from Paper~I ($V_a^{\rm Paper~I}$) has to be rescaled into 
$V_a^{\rm Paper~I,~Standard}$ ({\em i.e.} as the standard convention 
used in this work and others) through the relation 
\begin{equation}
V_a^{\rm Paper~I}=V_a^{\rm Paper~I,~Standard}
        \sqrt{\frac{6}{\delta(4-\delta^2)(4-\delta)}}\:.
\end{equation}

Second, the equation describing diffusion in energy has been 
modified and, the values of $K_0$, $V_c$ and $V_a$ that 
give the best fit to B/C data for a given $\delta$ must change at some 
level.
Notice that in Paper~I we used a source 
term corresponding to type (b) spectra (see Eq.~\ref{Typeb}), with $\gamma$ 
of each species that were set to their measured value (see details 
in Paper~I); this corresponds roughly to $\gamma \approx 2.7$ for all boron 
progenitors.

However, we find that the conclusions raised in Paper~I, in 
particular the behaviors reflected 
in Figs.~7 and~8 of Paper~I, are 
basically unchanged (it is not straightforward
to compare with present figures, but the careful reader can check
this result using the above scaling relation and the corresponding 
parameter combinations). To be more precise, it appears that
$K_0/L$ does not significantly change (for example, for $\delta=0.6$ and 
$L=2$~kpc, we still have 
$K_0/L\sim0.004$~kpc~Myr$^{-1}$, see Fig.~\ref{g28_beta_no_beta} left panel
-- this paper -- and Fig.~7 of Paper~I). As regards the galactic convective wind, 
$V_c$ is shifted towards higher values, whereas the $V_a/\sqrt{K_0}$ range 
remains roughly unchanged.

This can be easily understood: the additional term -- comparable
to a first order gain in energy, see Eq.~(\ref{reac1}) -- has to be balanced
to keep the fit good. This balance is ensured by enhanced adiabatic 
losses, {\em i.e.} bigger $V_c$. 
Other parameters are only very slightly affected by 
this new balance.

			
      \subsection{Jones et al. (2001), Moskalenko et al. (2002).}
Moskalenko et al. (2002) (hereafter Mos02)  use a description more refined than ours 
because they include a realistic gas distribution. Jones et al. (2001) 
(hereafter Jon01) take advantage of an equivalent description in terms of a leaky box 
formalism (use of a phenomenological diffusion coefficient)  -- for both 
wind model (no reacceleration) and minimal reacceleration model 
(no wind) -- to solve the diffusion equation.

Let us make a few comments at the qualitative level. First,
starting with Mos02's models, we can note that
convection has always been disfavored by these authors. For example,
in their first paper of a series (Strong \& Moskalenko, 1998), a gradient 
of convection greater than 7~km~s$^{-1}$~kpc$^{-1}$ was excluded.
We notice that this result was not very convincing since it is clear 
from an examination of their figures that none of the models they proposed 
gave good fits to B/C data. Thanks to many updates in their code, their fits were greatly
improved (Strong \& Moskalenko, 2001; Moskalenko et al., 2002) but if it
is now qualitatively good, it is hard to say how good it is since no 
quantitative criterion is furnished. Anyway,
our Fig.~\ref{g28_beta} allows us to understand why convection is disfavored in
such models. Actually, if $\delta\sim 0.3$ -- as in the Kolmogorov
diffusion slope hypothesis $\delta=1/3$ --, we see that for such a 
configuration, the best fits are obtained for $V_c\sim 0$~km~s$^{-1}$.

Similar comments apply to Jon01's models.
Given a Kolmogorov spectral index for the diffusion coefficient, 
their combined fit to B/C 
plus sub-Fe/Fe data is not entirely satisfactory. 
It improves for higher 
values of $\delta$ and in the convective model (they do not include
reacceleration in this model), their best fit being obtained for 
$\delta=0.74$. As the authors emphasized, the search in parameter space 
was not automated and they cannot guarantee that their best fit is 
the absolute best fit. 
Actually, the sub-Fe/Fe contribution to the $\chi^2$ value has to be taken
with care. First, the error bars are not estimated well enough to give a 
statistical meaning for $\chi^2$ values (see Sec.~\ref{quideu}) and a different 
weight should be considered for B/C and sub-Fe/Fe. 
Second, if the best parameters extracted 
from B/C data reproduce formally the same $\chi^2$ surface when applied
to the evaluation of sub-Fe/Fe (see Fig.~\ref{chi2_subfe_BC_values}), 
the direct search for the parameters minimizing $\chi^2$ for the same 
sub-Fe/Fe data gives constraints that are much weaker (see Fig.~\ref{chi2_subfe}). 
Thus, any conclusion including this ratio is from our
point of view far less robust. 
			     
       \subsection{Quantitative comparison, interpretation of $K_0$, $V_c$ and $V_a$}

			     
       \subsubsection{Justification of the differences between models}

Tabs.~\ref{comparaison2} and~\ref{comparaison2bis} give the results of Mos02 and 
Jon01 -- without any rescaling of any parameters --
compared to what is obtained here; only a few models are displayed.
\begin{table}[ht]
\caption{Diffusion parameters obtained in models with $\delta=0.30$, 
$\alpha=2.40$ for pure power-law source spectra.}
\label{comparaison2}
\begin{center}
\begin{tabular}{|c|c|c|c||c|c|c||c|c|}   \hline
$L$ & h&\ $\mu\times 10^{-3}$ &$h_{\rm reac}$& $K_0$ & $V_c$ & 
$V_a$ & $\chi^2_r$ & Ref. \\
(kpc) & (kpc) &(g~cm$^{-2}$) & (kpc) &(kpc$^2$~Myr$^{-1}$) & (km~s$^{-1}$) & (km~s$^{-1}$) & & \\\hline\hline
4. &$n(r)$ &1.6&4.0 &  $\sim0.201$ & 0.  &  30. & Good & Mos02$^\S$\\
3. & 0.2&2.4&1.0&  $\sim0.196$ & 0. & 40. & 1.8& Jon01$^\dagger$\\
3. &  0.1&1.0&0.1& $\sim0.0535$ & 0. & 105.8 & 4.2 & 
(Figs.~\ref{final_BC_chi2} and~\ref{final_BC_reste}, this paper)$^\ddagger$\\\hline
3. &  0.2 &2.4&1.0& $\sim 0.127$ & 0. & 47.3 & 4.4 & This work$^\ddagger$\\\hline
\end{tabular}
\end{center}\d
{~$^\S$ For this model, the exact values are $\delta=0.33$, $\alpha=2.43$.}\\
{~$^{\dagger}$ This model give best fit to flux using a slightly modified 
form for the source; $Q\propto R^{-2.40}/[1-(R/2)^{-2}]^{1/2}$.}\\
{~~$^\ddagger$ Corresponds to the best fit for the presented $L$, $\alpha$ 
and $\delta$ value.}
\end{table}

Tab.~\ref{comparaison2} shows $K_0$, $V_c$ and $V_a$ for $\delta=0.30$ and $\alpha=2.40$.
Taking the first three lines at face value, our values of $K_0$ and 
$V_a$ are very different from the others and our model seems to have 
a problem.
However, the matter disk properties (height and surface density) are different 
in these models. To be able to compare, we set these quantities to the 
values given in Jon01 and the resulting parameters are 
shown in the last line of Tab.~\ref{comparaison2}.

Actually, we know that in diffusion models,
the behavior is driven by the location of the closer edge, leading to a preferred
escape on this side. With $L=3$ kpc,
our three-dimensional model should behave as the two-dimensional
model with infinite extension in the $r$ direction of Jon01. 
This hypothesis can be validated if one takes their Eq.~(3.6). 
For the pure diffusion model (reacceleration and convection are discarded), one has a 
simple relation between $\mu$, $L$ and $K_0$ through an equivalent leaky box 
grammage
\begin{equation}
X_{dif}=\frac{\mu \;vL}{2K_0R^\delta}
\label{gram1}
\end{equation} 
A direct application of this result to our model with 
$h=100$~pc (third line of
Tab.~\ref{comparaison2}) using the scaling $\mu\rightarrow 2.4\times\mu$, leads
to a rescaling $K_0\rightarrow 2.4\times K_0$, consistent with results of the fourth 
line.

A similar expression may be obtained in the presence of galactic wind~:
in the wind model (their Eq.~4.6), one has
\begin{equation}
X_w=\frac{\mu \;v}{2V_c}\left[ 1-\exp\left(\frac{-V_c L}{K_0R^\delta}\right) \right]
\label{gram2}
\end{equation} 
Applied to the second line of Tab.~\ref{comparaison2bis}, this gives
$V_c\rightarrow 2.4 \times V_c$, leading in turn to $K_0\rightarrow 2.4\times K_0$,
also in very good agreement with the direct output of our code. 

\begin{table}[ht]
\caption{Diffusion parameters obtained in models $\delta=0.74$, $\alpha=2.35$.}
\label{comparaison2bis}
\begin{center}
\begin{tabular}{|c|c|c|c||c|c|c||c|c|}   \hline
$L$ & h&\ $\mu\times 10^{-3}$ &$h_{\rm reac}$& $K_0$ & $V_c$ & 
$V_a$ & $\chi^2_r$ & Ref. \\
(kpc) & (kpc) & (g~cm$^{-2}$) & (kpc) &(kpc$^2$~Myr$^{-1}$) & (km/s) & (km/s) & & \\\hline\hline
3. & 0.2&2.4&1.0&  $\sim0.024$   & 29.   & 0.    & 1.5 & Jon01 \\
3. &0.1&1.0&0.1&  $\sim0.0056$   & 15.5   & 35.3   & 3.0 &  This work$^\ddagger$\\\hline
3.  &0.2&2.4&0.1&  $\sim 0.0134$   & 36.5   & 26.5  & 3.1 &  -\\\hline
\end{tabular}
\end{center}
{~~$^\ddagger$ Corresponds to the best fit for the presented $L$, $\alpha$ 
and $\delta$ value.}
\end{table}

Even with this $\mu$ rescaling, the diffusion coefficients obtained 
by the different authors quoted above are still not fully compatible. 
Another possible effect, namely the spatial distribution of 
cosmic ray sources, is now investigated.
We note that in our model, the radial distribution of sources 
$q(r)$ follows the distribution of  supernov\ae\  and pulsar remnants. 
The choice of this distribution has an effect on B/C spectra and on the parameters giving 
the best fits. If we use a constant source 
distribution $q(r)=cte$ with $V_a$ set to 0 to follow Jon01, we find 
that $K_0$ is enhanced by about 10\%. 
We checked that it is also the case for results presented in Tab.~\ref{comparaison2}. 
Hence, it appears that results for $\delta=0.74$ of Jon01, though slightly 
different,  are not in conflict with ours. 
As regards $\delta=0.3$ and Mos2, using the scaling relation 
(\ref{gram1}) along with a 10\% decrease of $K_0$ for Jon01, we obtain
respectively $K_0=0.226$ (Mos02), $0.176$ (Jon01), $0.127$ (this paper)
kpc$^2$~Myr$^{-1}$.
Thus there is some difference between Mos02 and Jon01, which is not 
obvious when the values taken naively from Tab.~\ref{comparaison2} 
are compared. 
These discrepencies could have several origins: treatment of cross-sections 
(we checked that total and spallative cross sections -- taking into account 
ghost nuclei, see Paper I -- are compatible with recent data, 
{\em e.g.} Korejwo et al., 2002), 
average surface density in Mos02 that is probably not exactly $1.6$,
choice of data and fits for Jon01 that differ from ours (some of the
point they used are significantly lower than {\sc heao}-3's).
Finally, the fact that we scan the whole parameter space can make a 
difference from manual search. To conclude, results are qualitatively 
similar, but a few
quantitative differences remain. The intrinsic complications
and subtleties of the various propagation codes make it difficult to go 
further in the analysis of these differences.


       \subsubsection{Meaning of $K_0$}
\label{KetVc}

The normalization $K_0$ gives a measure of the efficiency of the 
diffusion process at a given energy. 
Its value can be predicted if (i) a good modelling of charged 
particles in a stochastic magnetic field and (ii) a good description 
of the actual spatial structure of this magnetic field, were available.
It is not the case and the precise value of $K_0$ is of little interest.
Moreover, the presence of effects other than pure diffusion can be 
mimicked, at least to some extent, by a change in $K_0$.
Eq.~(\ref{gram2}) gives a whole class of parameters giving the 
same results and can be used to extract an effective value of $K_0$ 
taking into account the effect of the size of the halo $L$ and wind 
$V_c$. This also explains the great range of values that can be found
in the literature.

This relation shows that there is also an indeterminacy of the absolute density of
the model, because as long as $h \times n_{\rm ISM}$ is constant, the grammage 
$X_{dif}$ is also constant.
Fortunately, a realistic distribution of gas can be deduced by more 
direct observational methods, so that a definite value of $n_{\rm ISM}$ 
can be used.


       \subsubsection{Galactic convective wind $V_c$}
\label{etVc}

We note that in our model, Galactic wind is perpendicular
to the disc plane and is constant with $z$. Actually, the exact form
of galactic winds is not known. From a self-consistent analytical description
including magnetohydrodynamic calculations of the galactic wind flow,
cosmic-ray pressure and the thermal gas in a rotating galaxy,
Ptuskin et al. (1997) (see also references therein) find a wind
increasing linearly with $z$ up to $z\sim 15$~kpc, with a $z=0$  
value of about 22.5 km~s$^{-1}$. 
Following a completely different approach,
Soutoul \& Ptuskin (2001) extract the velocity form able to reproduce
data from a one-dimensional diffusion/convection model. They obtain a decrease
from 35~km~s$^{-1}$ to 12~km~s$^{-1}$ for $z$ ranging from 40~pc to~1~kpc 
followed by an increase to~20~km~s$^{-1}$ at about 3~kpc. 
For reference, our values for the best fits correspond to about 15~km~s$^{-1}$.
The difficulty to compare constant wind values to other $z$-dependences
is related to the fact that cosmic rays do not spend the same amount of time 
at all $z$, so that there cannot be a simple correspondence 
(see also next section) from one model to another.
As a result, all the above-mentioned models are formally different,
with different inputs (spectral index, diffusion slope).
Nevertheless, their values are roughly compatible, Ptuskin et al's model
providing the grounds for a physical motivation for this wind.
However, an even more complicated form 
of the Galactic wind could be relevant for a global description
of the Galaxy (see Breitschwerdt et al., 2002).


       \subsubsection{Interpretation of the Alfv\'enic speed $V_a$}
\label{ouinde}
Above, we gave some elements to compare $V_a$ values from various works.
Actually, secondary to primary ratios are not determined directly by the
Alfv\'en speed in the interstellar medium, but rather by an effective value:
\begin{equation}
   V_a^{\rm true}=\sqrt{\frac{h_{\rm reac}}{h}}\times \frac{V_a^{\rm
   eff}}{\sqrt{\omega}}
   \label{trueVa}
\end{equation}
First, the parameter $\omega$ characterizes the level of turbulence and is often
set to 1 (Seo \& Ptuskin, 1994). Our model, as others, uses
\begin{eqnarray}
     \omega(z)= \left\{
      \begin{tabular}{cl}
      $1$ & if $z<h_{\rm reac}$,\\
      $0$ & otherwise;
      \end{tabular}
     \right.
   \end{eqnarray}
as a crude approximation of the more complex reality. 

Second, the total rate of reacceleration (at least in a first approximation, 
see discussion below) is given by a convolution of the time spend in the 
reacceleration zone and the corresponding {\em true} Alfv\'en speed in this zone. 
There is a direct analogy with the case of spallations and the determination 
of the true density in the disc, as discussed above.
The problem is still somewhat different, as
there are no direct observational clues about the size of the reacceleration
zone, or said differently, about $\omega(z)$. 
This leads to a degeneracy in $h/h_{\rm reac}$ that holds as far as 
$h_{\rm reac}\ll L$, due 
to the structure of the equations in the thin disc approximation. 
For example, a model such as Strong et al's
that uses $h_{\rm reac}=L$ cannot be simply scaled to ours. 
A cosmic ray undergoing reacceleration at a certain height $z$ has a 
finite probability of escaping before it reaches Earth, this probability 
being greater for greater $z$. 
As a result, the total reacceleration undergone by a cosmic ray is actually not 
a simple convolution of the reacceleration zone times 
the Alfv\'en speed in this zone, but rather should be an average
along $z$ taking into account the above-mentioned probability (in principle, this
remark also holds for the gaseous disk, though the latter is known to be very thin,
$\sim$ a few hundreds of pc).

To conclude, there are basically three steps associated with three levels 
of approximations to go from the $V_a$ deduced from cosmic ray analysis 
to the physical quantity.  
First, if $\omega(z)$ is approximatively constant with $z$, how large is the 
reacceleration zone height? The second level is related to the possibility that 
$\omega(z)$ strongly depends on $z$ in a large reacceleration zone. If it is too 
large, the link with the phenomenologically  equivalent quantity in a thin zone is 
related to the vertical occupation of cosmic rays. However, this latter
possibility seems to be unfavoured by {\sc mhd} simulations (see Ptuskin et
al., 1997).
Finally, with the above parallel between interpretation of $\mu$ and $V_a$, we see 
how misleading it is to obtain precise physical quantities from our simple model,
since there is no one-to-one correspondence between reality and simplified models. 
This discussion shows that even if the actual 
derived Alfv\'en speeds are consistent with what is expected from ``direct" observation
($\sim 10-30$~km~s$^{-1}$), the cosmic ray studies would certainly
be not very helpful in providing physical quantities better than a factor of two. 
If we reverse the reasoning and retain our best models with $L=6$ kpc,
we could conclude that $\sqrt{h/h_{\rm reac}}$ must be $\sim 4$ in 
order to give realistic values for $V_a$ (with evident {\em a priori} about $\omega(z)$).


\subsubsection{The evolution of propagation models}
\label{critique}
As suggested by the previous discussion, there are several propagation
schemes, 
each associated with numerous configurations,  that are able to
explain the B/C data. 
Thus, the discussion should not be about the correctness
of all these models (leaky boxes, two or three-dimensional diffusion models
and their inner degeneracy), but rather about their domain of validity.
As a matter of fact, they are all equivalent, as far as stable cosmic rays
around GeV/nuc energies are considered.

Starting with the leaky box; it has been shown more than thirty years ago
(Jones, 1970) that the concept of ``leakage-lifetime" was appropriate for 
the charged nuclei considered here (see also Jones et al., 1989), 
even if it broke down for $e^-$ (all orders in the development in ``leakage" eigenmodes 
contribute because of synchrotron or inverse Compton losses) and for radioactive nuclei 
(Prishchep \& Ptuskin, 1975). 
The leaky box, due to its simplicity,
is very well suited for the extraction of source abundances (elemental as well 
as isotopic). 
It can also
be used for secondary antiproton production, since the same processes
as for secondary stable nuclei are at work. However, as emphasized in 
Paper~II,
leaky box models are not able to predict any primary contribution in the antiproton
signal, since it requires the knowledge of the spatial distribution
of primary progenitors. Considering a possible extension of leaky box models
for stable charged nuclei to high energy  ($\sim$~PeV), it has been demonstrated 
in Maurin et al. (2002) that they are to a good approximation sufficient to describe the
evolution of cosmic rays. Last, it is well known that leaky box parameters are just
phenomenological with only a distant connection to physical quantities.

This was further realized by Jones (1978, 1979) who first remarked that 
the phenomenological behavior of the escape length at low energy could 
be due to the presence of  a Galactic wind.
Jones et al. (2001) investigated further this idea and generated several 
equivalent phenomenological escape lengths from 
several possible physical configurations of a one-dimensional diffusion model. 
The relation between one-dimensional models and leaky box models is thus firmly 
established and very well understood. Moreover, this relation 
elucidates some of the physical content of leaky box models. 
Now if one wishes to overcome the 
inherent limitations of these models and say, to compute some primary antiproton 
component, one has to go through a three-dimensional model. 
It is likely that these models can also be related to the Jones et al. models
(see Taillet \& Maurin, in preparation). Several arguments used in the previous sections
illustrate this view, but this occurs at least if the halo size is small
compared to radial extension of the Galaxy.

In the semi-analytical two-zone model used here, it is possible to evaluate 
the primary antiproton component (see Barrau et al., 2001) and to take 
into account radioactive species, even in the presence of a local very 
underdense bubble (see Donato et al., 2002 for details).
Our model fails to consider
species such as $e^-$ and $e^+$, since the latter suffer from large energetic losses
in the halo so that no simple semi-analytical approach can be used.
The parameters extracted from these models are much easier to interpret 
in terms of physical quantities.

Most of the limitations mentionned above are overcome by Strong et al's models.
In this fully numerical model, all cosmic ray species can be computed self-consistently
with the same propagation parameters. 
The main difference with our model is that a more realistic matter 
distribution is used instead of a thin homogeneous disk.
They also consider that reacceleration occurs in the whole diffusion 
halo, which in our opinion is an approximation no more justified than the 
fact to confine it in a thin disk (see discussion in Sec.~\ref{ouinde}). 
Considering the gas distribution, both models are equally predictive for 
the charged nuclei (including antiprotons, see Fig.~9 of Paper~II). 
On the one hand, our approach is better suited to scan the whole parameter space
as we did in Paper~I and in this paper. 
On the other hand, the Strong et al. models can check the consistency of $e^-$, 
$e^+$ and $\gamma$ with observations, 
and can include whatever deviation from ideal cases for $K_0$, $V_a$, 
$V_c$ and more generally for any ingredient that enters in the description of 
propagation models.

To conclude about models and their use, Jones et al.'s approach is probably the best and 
simpliest way to understand how physical parameters affect the propagated flux.
Our model is very well suited for a consistent evaluation of all charged nuclei
and extraction of propagation parameters; furthermore it is an intermediate 
step where general behaviors can still be analytically explored 
(Taillet \& Maurin, in preparation). In the Strong et al. model, all fine effects can be studied
and modelled, with the counterpart that the numerical approach makes 
the physical intuition of the results less straightforward.
In its present form, Strong et al's model can be viewed
as a fully numerical version of ours, so that their behaviors are very close.
This discussion could leave the reader with a feeling that apart from 
these different modellings left to personal taste,
galactic propagation phenomena are well understood.
It is surely not the case! 
Even if all these models are equivalent to describe 
the {\em local} observations of charged cosmic rays, they lead to 
very different conclusions and interpretations when the spatial variation 
of the cosmic ray density is considered.
As an illustration of the poor current understanding of this global 
aspect, we mention the ever-lasting problem of 
the gamma ray excess about 1~GeV towards the Galactic center or 
the too flat radial $\gamma$-ray distribution observed in the disc
(see Breitschwerdt et al., 2002).

\section{Conclusion}
Forgetting for a while some of our theoretical {\em a priori} about
the diffusion power spectrum, a new picture of cosmic ray propagation 
seems to emerge, motivated by the B/C analysis. 
In this new picture, high values for the diffusion coefficient spectral index 
($\delta\gtrsim 0.6-0.7$) and source spectral indices $\alpha \sim 2.0$ are favored.  
This latter result is rather satisfactory: as emphasized
in a recent working group report on SNR shocks (Drury et al., 2001), 
even {\em if nonlinear acceleration models do 
not produce precise power-law spectra [...] the effective differential energy 
spectral index is close to 2.0.
Furthermore, as pointed out in a series of papers by Vainio and Schlickeiser 
(2001, and see references therein), diffusive shock wave acceleration 
naturally yields smaller values of $\alpha $ if the correct scattering center 
compression ratio is used instead of the gas compression ratio."\\}
This trend should be carefully
analysed and discussed in the light of measured differential fluxes, in order
to confirm or point out the possible inconsistencies in the current propagation treatments
(see companion paper, Donato, Maurin \& Taillet, in preparation). Briefly, the major 
arguments against large $\delta$ come
from anisotropy measurements at high energy and from theoretical 
preference for
Kolmogorov-like turbulence spectra. However, Ptuskin et al. (1997) -- in their self-consistent analytical 
propagation model including gas, cosmic ray and magnetic field -- derived $\delta\sim 0.55$, 
$\alpha\sim 2.1$ and argue that the observed anisotropy could be as well due 
to a particularity of the local structure of the Galactic magnetic field. 
Theoretical objections against too high values of $\delta$
are probably more robust.

For the rest, the conclusions of this paper can be summarized as 
follows : (i) we performed for the first time 
a full analysis of diffusion/convection/reacceleration models in the whole
6-dimensional parameter space ($\alpha$, $\delta$, $K_0$, $L$, $V_a$, $V_c$),
and $\delta\sim 0.7-0.9$; the values $\alpha\sim2.0$ are preferred; (ii) 
this preference holds whatever the specific form of the spectrum at low energy;
the numerical values of the other parameters are also only slightly modified 
by this low energy dependence even though deviation from a power-law
at low energy is preferred. 
The study of fluxes should give a more definite answer;
(iii) $K_0$ scales logarithmically with $\delta$ and models with small halos
tend to one-dimensional models with a simple relation between $\mu$, 
$K_0$, $L$ and $V_c$ (see also Taillet \& Maurin, in preparation); 
(iv) several existing models are 
compared and the qualitative and quantitative differences 
between them are studied and partially explained.

\section*{Acknowledgments}
D.M. would like to thank Aim\'e Soutoul for valuable remarks and
Michel Cass\'e and Elisabeth Vangioni-Flam for many interesting discussions. 
We thank Prof. Schlickeiser for providing us with the diffusion 
coefficients in the different schemes in a useful form.

\end{document}